\documentclass[11pt]{article}
\usepackage{arxiv}
\usepackage{graphicx}%
\usepackage{multirow}%
\usepackage{amsmath,amssymb,amsfonts}%
\usepackage{xcolor}%
\usepackage{soul}

\usepackage{subcaption}
\usepackage{adjustbox}
\usepackage{rotating}

\usepackage{booktabs}
\usepackage{tabularx}
\usepackage[hidelinks]{hyperref}
\usepackage{xurl}
\usepackage[group-separator={,}]{siunitx}

\usepackage{glossaries}

\newacronym{GIS}{GIS}{Geographic Information System}
\newacronym{GPS}{GPS}{Global Positioning System}
\newacronym{GTFS}{GTFS}{General Transit Feed Specification }
\newacronym{NAICS}{NAICS}{North American Industry Classification System}
\newacronym{ODbL}{ODbL}{Open Database License}
\newacronym{OLS}{OLS}{Ordinary Least Squares}
\newacronym{OSM}{OSM}{OpenStreetMap}
\newacronym{POI}{POI}{Point of Interest}
\newacronym{TPDD}{TPDD}{Temporal Population Density Data}
\newacronym{WGS84}{WGS84}{World Geodetic System}

\usepackage[style=numeric,sorting=none,defernumbers=true]{biblatex}
\usepackage{authblk}

\usepackage{orcidlink}

\begin{document}

\title{Public transport in the 15-minute city}

\author[1,*]{{Zsófia } {Zádor} \orcidlink{0000-0003-3685-4748}}

\author[2,*]{{Gergő} {Pintér} \orcidlink{0000-0003-4731-3816}}

\author[3,*]{{Máté} {Mizsák}}

\author[4]{{Bence} {Kovács}}

\author[5]{{Imre} {Felde} \orcidlink{0000-0003-4126-2480}}

\author[2,3]{{Balázs} {Lengyel} \orcidlink{0000-0001-5196-5599}}

\affil[1]{{Northeastern University London}, {{EC1V 0HB} {London}, {United Kingdom}}}

\affil[2]{{Corvinus University of Budapest}, {{1093} {Budapest}, {Hungary}}}

\affil[3]{{Agglomeration, Networks, and Innovation Lendület Research Group}, {Centre for Economic and Regional Studies}, {{1097} {Budapest}, {Hungary}}}

\affil[4]{{Data Bank}, {Centre for Economic and Regional Studies}, {{1097} {Budapest}, {Hungary}}}

\affil[5]{{Óbuda University}, {{1034} {Budapest}, {Hungary}}}

\affil[*]{These authors contributed equally to this work.}

\maketitle
\begin{abstract}
The 15-minute city is a powerful planning concept to counter car-dependence by promoting active mobility to amenities and fostering inclusive urban environments. However, this policy has challenges in amenity-poor urban peripheries. Public transport remains underexplored in this discourse despite its role in distant access.
Here, we propose a framework that incorporates public transport into the 15-minute city model using openly available data.
By comparing Helsinki, Madrid, and Budapest, we demonstrate that multimodal mobility substantially increases access to amenities and enhances socio-spatial integration within a 15-minute reach. Although urban periphery benefit significantly from radial or high-speed public transport lines in their social mixing potential, such lines alone do not improve their access to amenities. These findings underscore the need to optimize polycentric public transport networks that can improve inclusive urban accessibility and complement active mobility in polycentric cities.
\end{abstract}

\begin{refsection}

\section{Introduction}
\label{sec:introduction}

The COVID-19 pandemic significantly accelerated the adoption of New Urbanism, an urban planning philosophy advocating for compact, mixed-use, and walkable neighborhoods \cite{Duany1994, Montgomery2013}. Coupled with growing climate change concerns, this shift has promoted active mobility and the idea of localized access to amenities within short distances \cite{Burton2000} and  is epitomized by the concept of the 15-minute city, which aims to revitalize urban neighborhoods. The goals of the 15-minute city extend beyond merely reducing traffic to include improving urban health and strengthening inclusive local communities \cite{Allam2022a, Moreno2021, Logan2022, pozoukidou202115}. The increasing relevance of this concept in urban planning has spurred extensive empirical research examining its feasibility and potential drawbacks (e.g. \cite{abbiasov202415, GraellsGarrido2021, Zhang2023}). 

A notable challenge lies in implementing proximity-driven development in urban peripheries that often lack essential amenities \cite{bruno2024universal, OECD2024cities} and have experienced the greatest growth in low-income populations \cite{berube2006two, liu2021suburbanization, allard2017places}. This raises a key question: how can access to amenities be made sustainable in remote neighborhoods \cite{poorthuis2023moving, colacco2025does}? 

Public transport offers a way to extend the reach of sustainable local living, improving access to daily essentials, while supporting inclusive cities \cite{Moreno2021}. Mixed modes of mobility, which typically combine active travel with community transit, are a well-established approach to reducing car traffic \cite{Marshall2000}. Nevertheless, their potential to complement active mobility in the 15-minute city context, particularly to improve access to amenities and to diverse social groups, has been overlooked.

In this paper, we develop a 15-minute city framework that incorporates public transport with walkable neighborhoods, enabling amenity access and more inclusive community life in residential areas. 
To this aim, we extend the traditional 15-minute walking distance to a multimodal mobility approach, which is an emerging area in urban data science \cite{Alessandretti2016, Alessandretti2017, Alessandretti2023}.
In this approach, we measure the access to amenities and to people from different socio-economic groups. 

We focus on three European cities, Helsinki, Madrid and Budapest, where public transport is accessible within 10-minute walk for around 90\% of the total urban population \cite{OECD2022}. Yet, this selection provides diverse contexts to investigate the research question: while the public transport in Madrid excels in linking the population to urban opportunities, this access is around OECD average in Budapest \cite{OECD2024cities}. Furthermore, by including Helsinki, where public transport provides good access, we can analyze the role of transit in a coastal city as well. 

The empirical exercise mainly leverages openly available data. First, we collect General Transit Feed Specification (GTFS) data that contains the time tables of the public transportation providers in each city. Second, we use OpenStreetMap (OSM) to locate and categorize amenities. Finally, we leverage publicly available data from Madrid and Helsinki that captures the socio-economic status of residents aggregated to a fine spatial scale. Such data is not available in Budapest, but we can use real estate prices that are good indicators of the status of neighborhoods \cite{Juhasz2023}. In addition, we have access to temporal population density data in Budapest that contains the number of individuals who are present in a given hour in high-resolution grids stratified by socio-economic groups. This information allows us to understand how public transport can help social mixing within a 15-minute range by comparing residential and experienced segregation \cite{Athey2021}. All data sets are explained in detail in Materials and Methods.

Using the GTFS time tables, we first extract the public transport stops and cluster them into groups of stops where transit between lines is possible. Then, we generate a network between the stop clusters, in which the edge weight represents the average travel time. By setting the transit     time, this network enables us to identify the area of multimodal mobility access that can be reached within 15 minutes, combining 5-minute walk to the station with 10 minutes of travel. Then, we can quantify the extra access to amenities and extra opportunities for social mixing the multimodal mobility provides compared to areas defined by 15 minutes of walking. Finally, we predict access and mixing premium of multimodal mobility with topological and socio-demographic characteristics of accessible areas.

We find that multimodal mobility increases the number of accessible amenity types and enhances opportunities for social mixing in all three cities. These improvements are greater in amenity-poor and segregated neighborhoods but are negatively correlated with distance from the city center. Even with multimodal mobility, social-mixing varies across socio-economic groups, with different classes experiencing different levels of connectivity. Yet, we find that the extent of multimodal reach to more amenities is important for the rich and poor in all cities. However, in urban peripheries, accessible areas often take on an elliptical shape due to the dominance of high-speed transit lines, which primarily connect to the city center. Our findings suggest that such elliptic multimodal access is not efficient in providing access to more types of amenities in urban peripheries, while it can improve their social mixing potential. 

In sum, this paper establishes a new empirical framework to study how public transport and multimodal mobility can enhance urban liveability, reduce car dependence, and expand access across socio-economic groups in 15-minute cities. The results imply that proximity-driven developments that focus on accessibility to amenities in urban peripheries should be complemented by polycentric public transport.



\section{Results}
\label{sec:results}

\begin{figure}[t!]
\centering
\includegraphics[width=\textwidth]{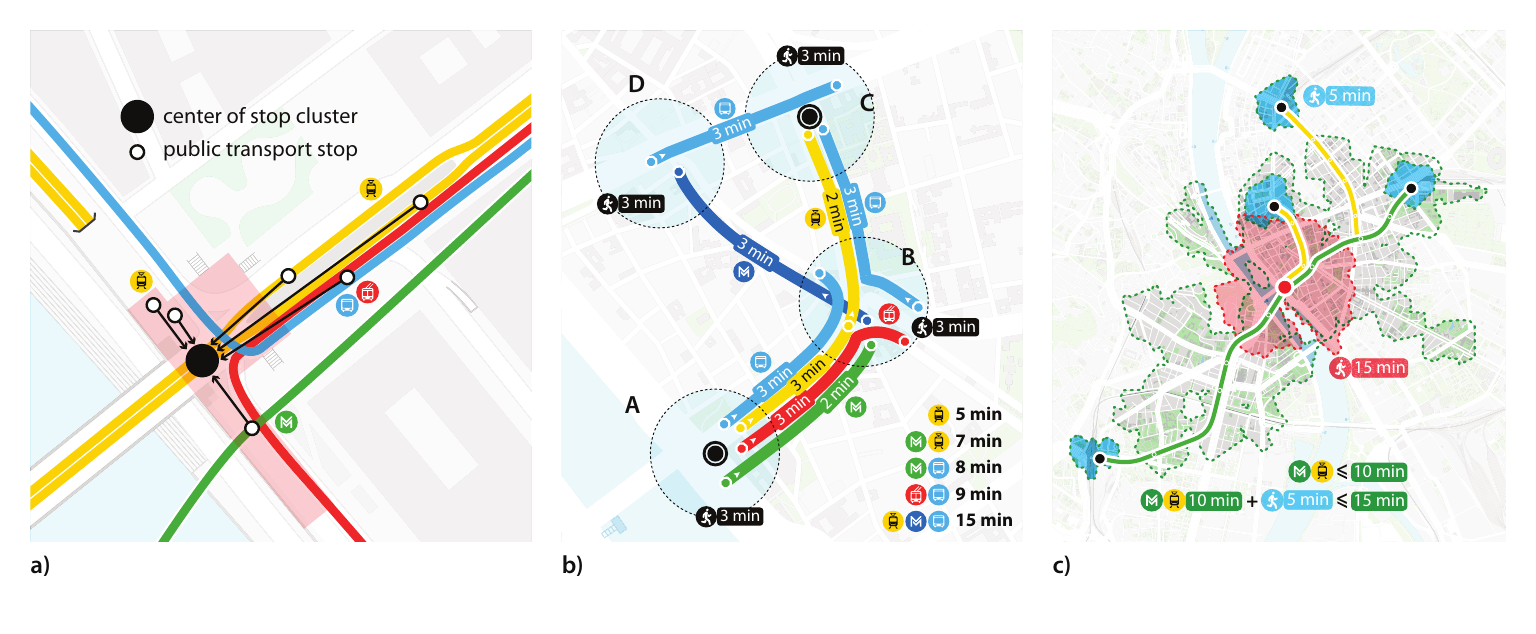}
\caption{\textbf{Creation of the accessibility network, using public transport time-table data.} \textbf{a)} Public transport stops are clustered into mobility hubs. The figure example is "Fővám tér" in Budapest that represents a collection of bus, tram, metro and trolley bus stops. \textbf{b)} Considering all lines and their time tables, we create a multilayer network between mobility hubs weighted by travel times and identify the fastest public transport access assuming 3-minute transit between lines. Here, we consider trips between Fővám tér (A) and Astoria (C). \textbf{c)} We depict the area of 15-minutes walk from the mobility hub on the street network and identify the maximum spread of multimodal access that combines 10 minutes of public transport travel --using the fastest rides-- that can be reached with 5 minutes walk.}
\label{fig1}
\end{figure}

To compare 15-minute walk accessibility with combined walking and public transport access, we first generate a multilayer network that connects clustered public transport stops with travel time among them (Figure \ref{fig1}). The clusters of stops represent squares or major junctions that can function as transit between transport lines; we refer to these as mobility hubs (Figure \ref{fig1}a). We use time table data, explained in Section \nameref{sec:materials_and_methods}, to construct the edges of the multilayer network between mobility hubs that contain all transport links with their travel times. Following consultation with practitioners, we assume that transit between lines can happen in 3 minutes, and that onward vehicles depart immediately upon a passenger's arrival. This enables us to calculate an estimated travel time between all hubs and identify the shortest path from the set of all possible links between any hub pairs. Figure \ref{fig1}b illustrates the example of potential trips between Fővám tér (A) and Astoria (C), two central mobility hubs in Budapest. The quickest access is using a tram that directly connects A and C (yellow line) but one could use various other lines and transit at Kálvin tér (B) and then even at Frenciek tere (D). Detailed explanation of the clustering procedure and network generation in the Supplementary Information (SI1).

Based on the shortest paths in the multilayer public transport network, we then delineate the accessible areas from each mobility hub combining 10 minutes of transport ride using any lines and 5 minutes walking (Figure \ref{fig1}c). Finally, we compare this multimodal access to an area covered by a walking radius of 15-minutes in the street network. In sum, this approach yields a ''15-minute walk'' polygon and a ''multimodal access'' polygon that are the basis of our further analyses. We will compare access to amenities and socio-economic groups across these polygons and explain additional accessibility using public transport by the size and shape of multimodal polygons and their location in the city.

The first characteristic of multimodal access premium is the size of the additional area $S_p=S_{multimod}-S_{walk}$ that one can cover with multimodal mobility $S_{multimod}$ (union of green and pink areas in Figure \ref{fig1}c) compared to the walking area $S_{walk}$ (pink area in Figure \ref{fig1}c). There is an intuitive positive relationship between $S_p$ and the diversity of accessible amenities and socio-economic groups. Yet, the intensity of this relationship might differ across cities and neighborhoods; thus, investigating this heterogeneity can inform us about the role of public transport in 15-minute access in specific locations.

\begin{figure}[ht]
    \centering
    \begin{subfigure}{0.84\linewidth}
        \includegraphics[width=\linewidth]{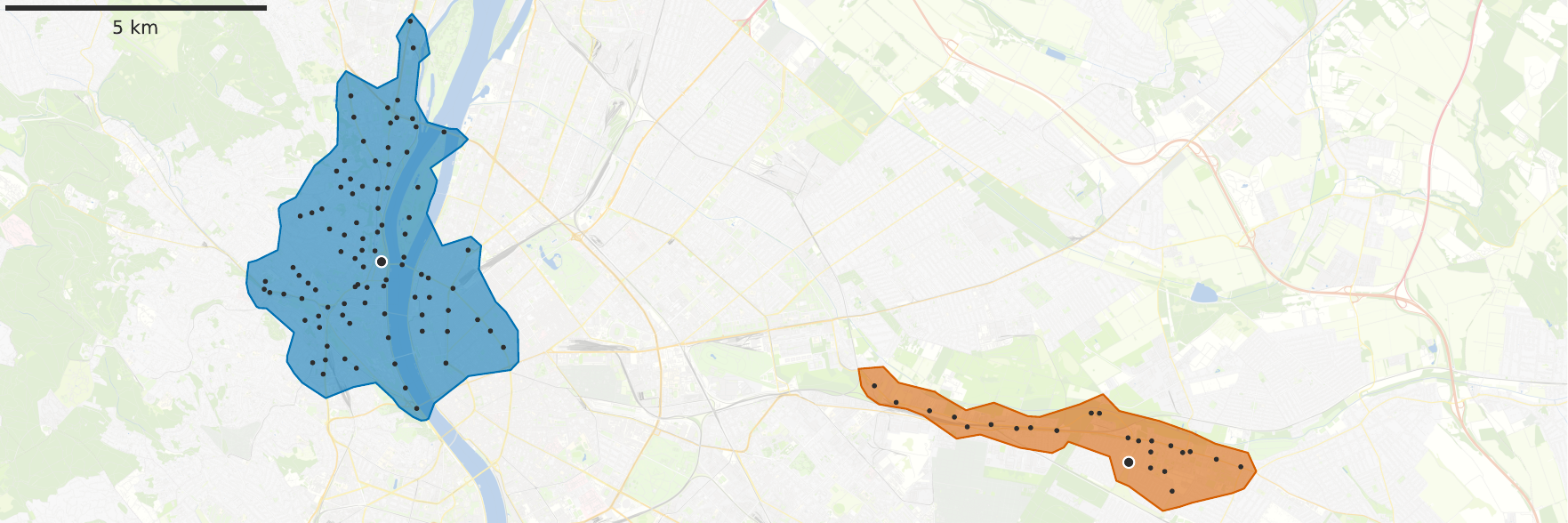}
        \caption{}
        \label{fig:ellipticity_examples}
    \end{subfigure}
    \hfill
    \begin{subfigure}{0.28\linewidth}
        \includegraphics[width=\linewidth]{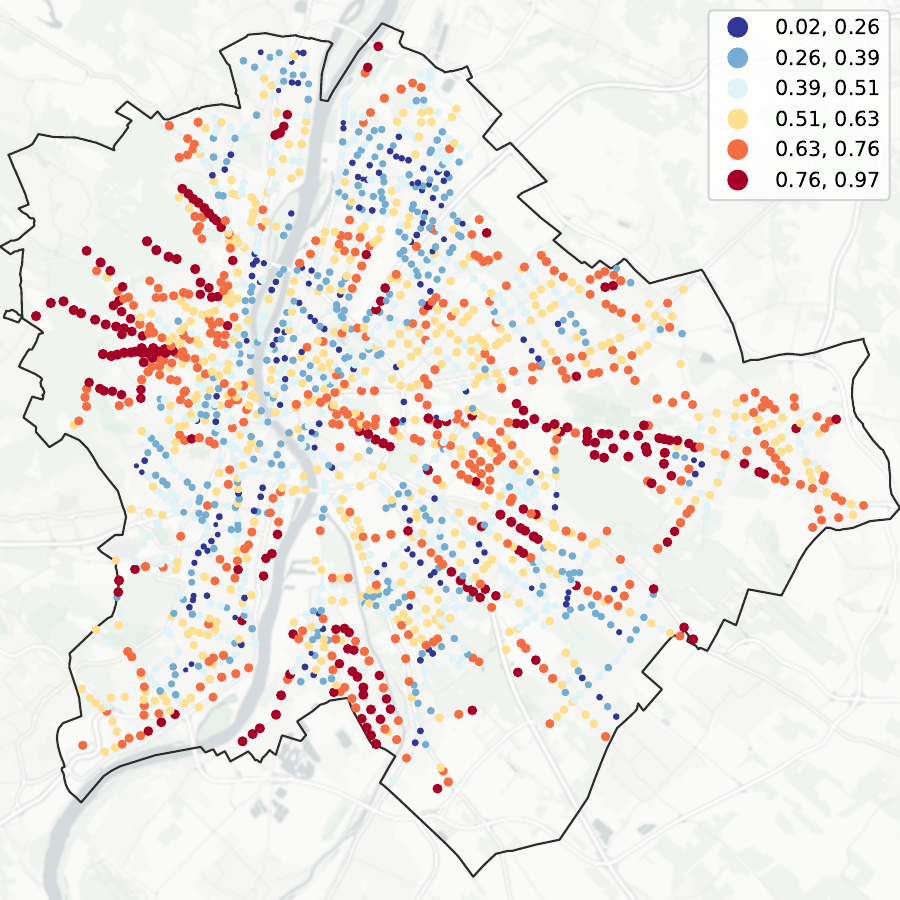}
        \caption{}
        \label{fig:budapest_ellipticity}
    \end{subfigure}
    \begin{subfigure}{0.28\linewidth}
        \includegraphics[width=\linewidth]{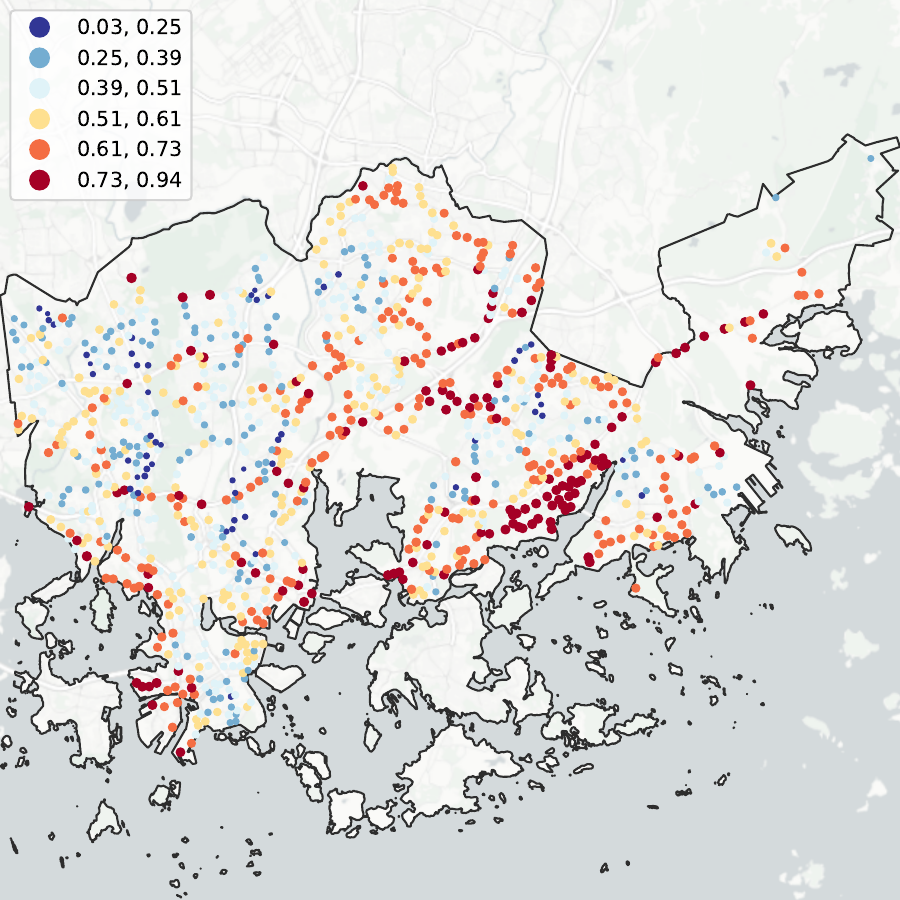}
        \caption{}
        \label{fig:helsinki_ellipticity}
    \end{subfigure}
    \begin{subfigure}{0.28\linewidth}
        \includegraphics[width=\linewidth]{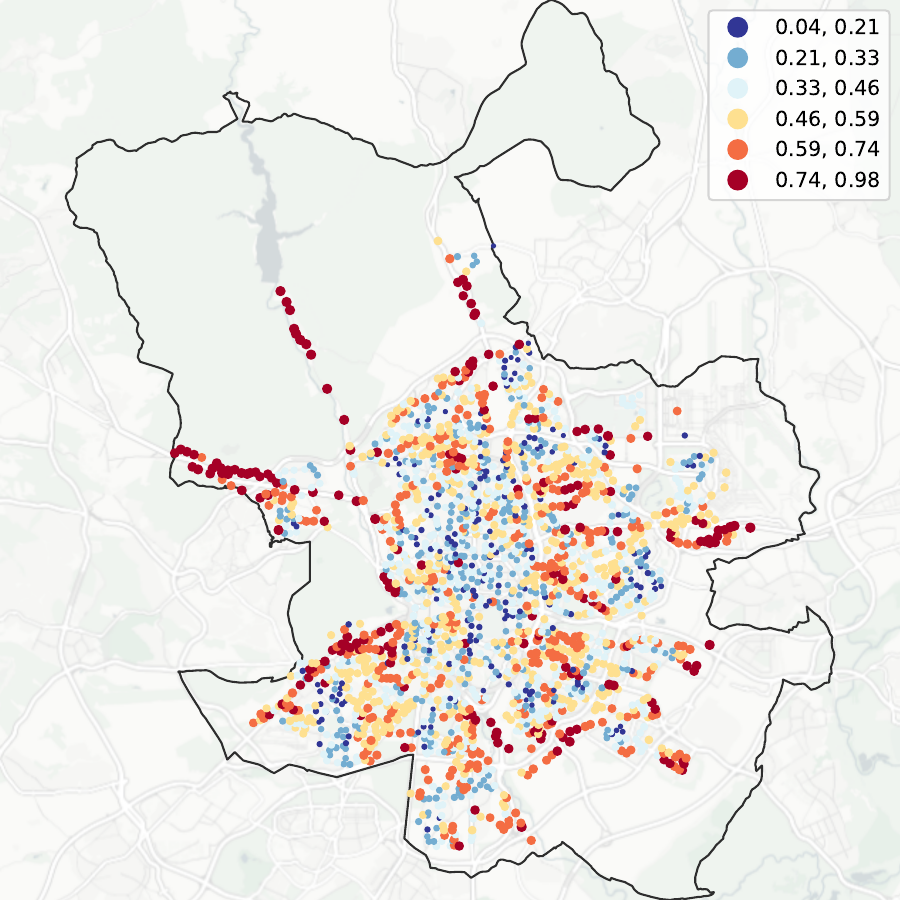}
        \caption{}
        \label{fig:madrid_ellipticity}
    \end{subfigure}
    \caption{\textbf{Ellipticity of multimodal access $E_p$.} \textbf{a)} Multimodality can extend access evenly (creating a round shape as the blue area) and non-evenly (creating an elliptic, leaf shape such as the orange area). High $E_p$ values define access corridors in urban peripheries in \textbf{b)} Budapest, \textbf{c)} Helsinki, and \textbf{d)} Madrid.}
    \label{fig:fig2}
\end{figure}

Next, we characterize the shape of the multimodal access polygons (Figure \ref{fig:fig2}). As one can see in Figure \ref{fig:fig2}a, public transport can extend access to multiple directions in a relatively balanced way (for example, the purple area in Figure \ref{fig:fig2}a), but can also extend to a single direction unevenly (see the orange area in Figure \ref{fig:fig2}a). This topological characteristic is important to include in the analysis, since multimodal expansion to multiple directions versus a single direction makes a qualitative difference in public transport strategies. To characterize the evenness of directions, we define the ellipticity of the multimodal access as
$E_p = 1 - \frac{b_p}{a_p},$
where \(a_p\) and \(b_p\) are the semi-major and semi-minor axes of the ellipse with the same second moments as polygon \(p\). The metric is explained in detail in Section \nameref{sec:materials_and_methods}.

Peripheral neighborhoods exhibit statistically significantly higher values of $E_p$ --that are more than median distance from the center-- than in more central places (Welch's t-test, $p < 0.01$ in all cases), though effect sizes are small (Cohen's d = -0.13 to -0.16, with average ellipticity values in Budapest: 0.507 vs. 0.531; Helsinki: 0.531 vs. 0.560; Madrid: 0.456 vs. 0.482). Indeed, high values of $E_p$ define corridors towards the center in the outskirts of Budapest ((Figure \ref{fig:fig2}b)), and Madrid ((Figure \ref{fig:fig2}d)), while the values are high along the seaside in Helsinki, as expected from a coastal city (Figure \ref{fig:fig2}c). Thus, we will use $E_p$ to understand the role of corridor-type public transport in providing 15-minute access in urban peripheries.

\begin{figure}[ht]
    \centering
    \begin{subfigure}{0.4\linewidth}
        \includegraphics[width=\linewidth]{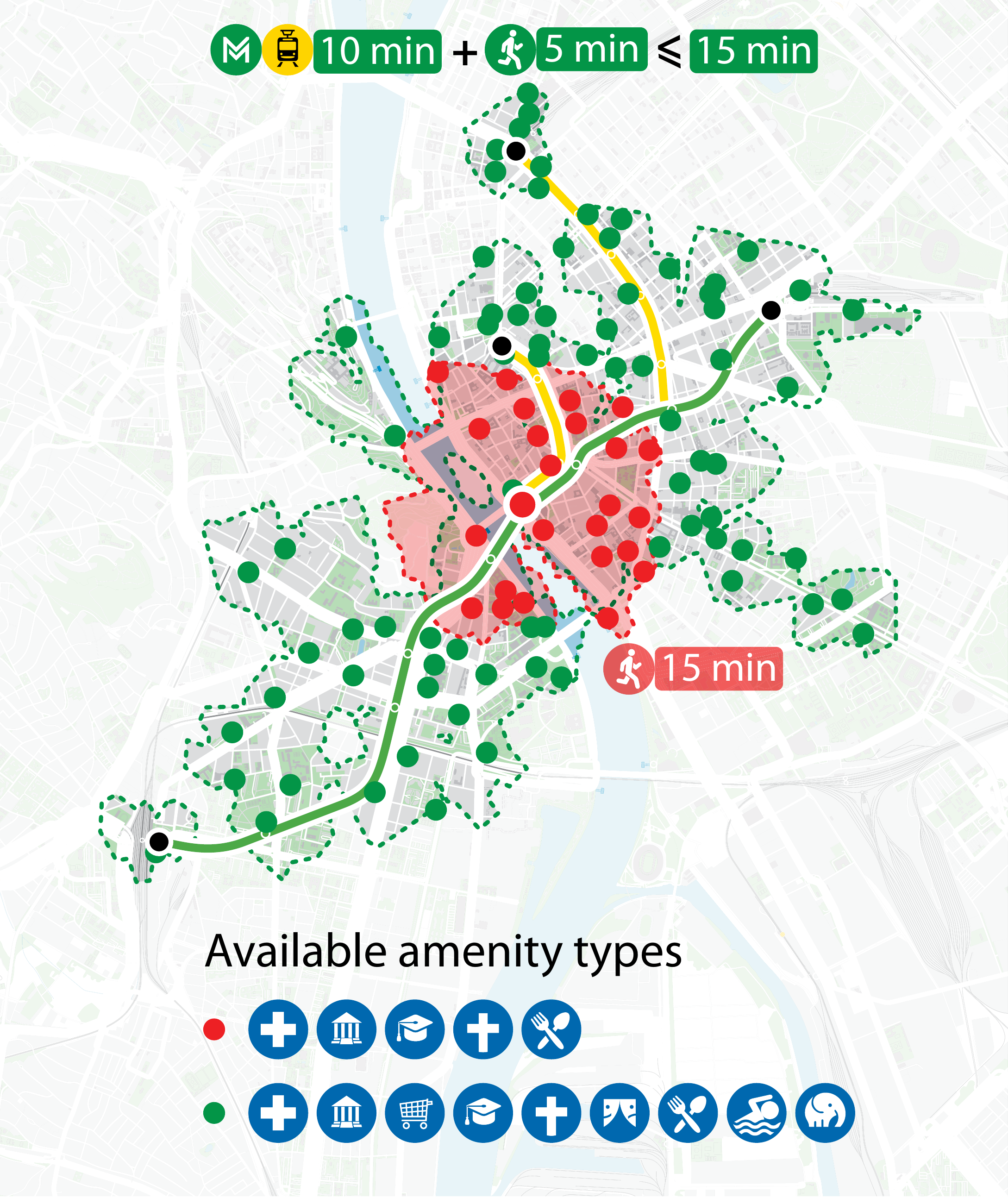}
        \caption{}
    \end{subfigure}
    \begin{subfigure}{0.4\linewidth}
        \includegraphics[width=\linewidth]{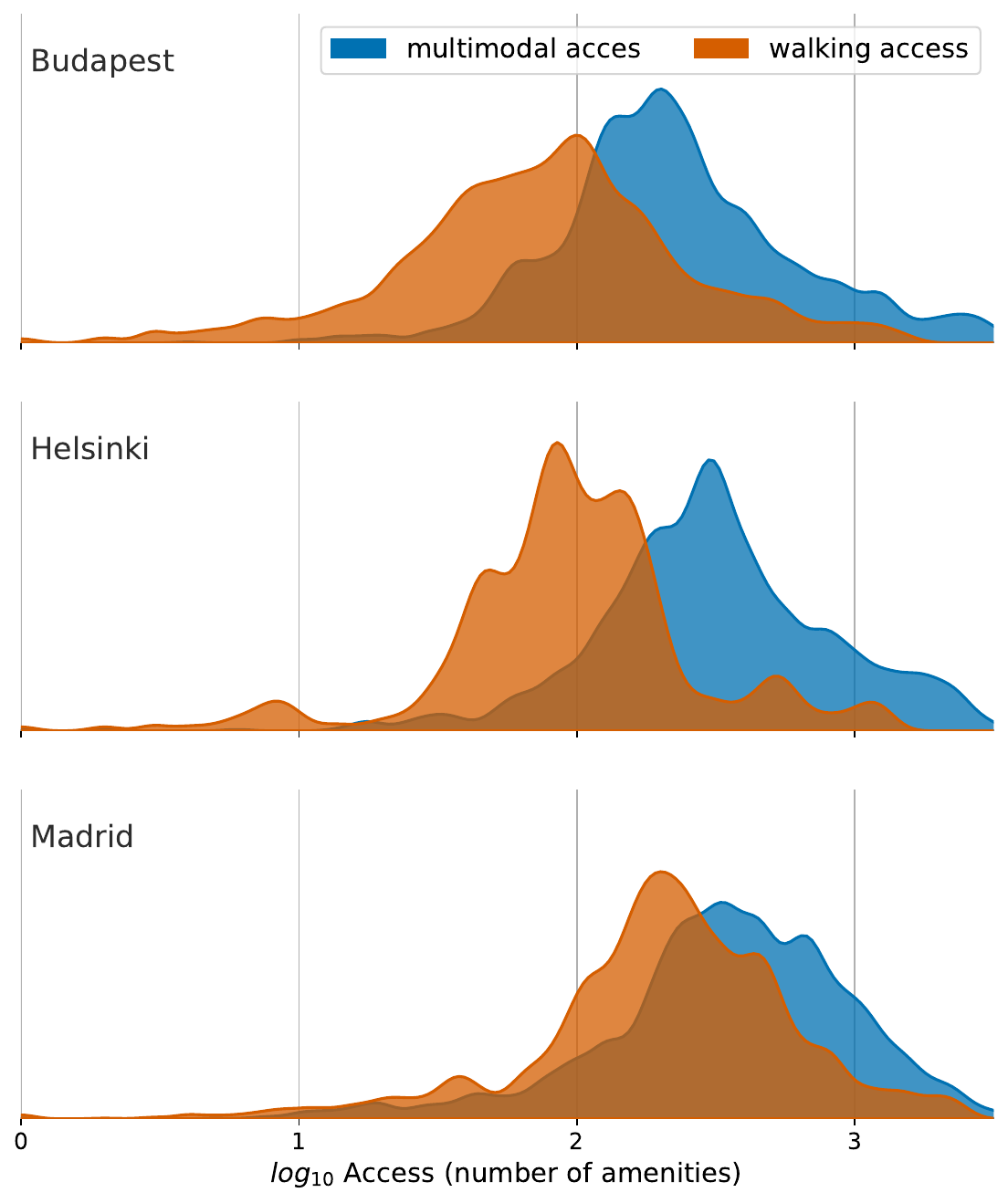}
        \caption{}
    \end{subfigure}

    \begin{subfigure}{0.8\linewidth}
        \includegraphics[width=\linewidth, height=0.25\textheight, keepaspectratio=false]{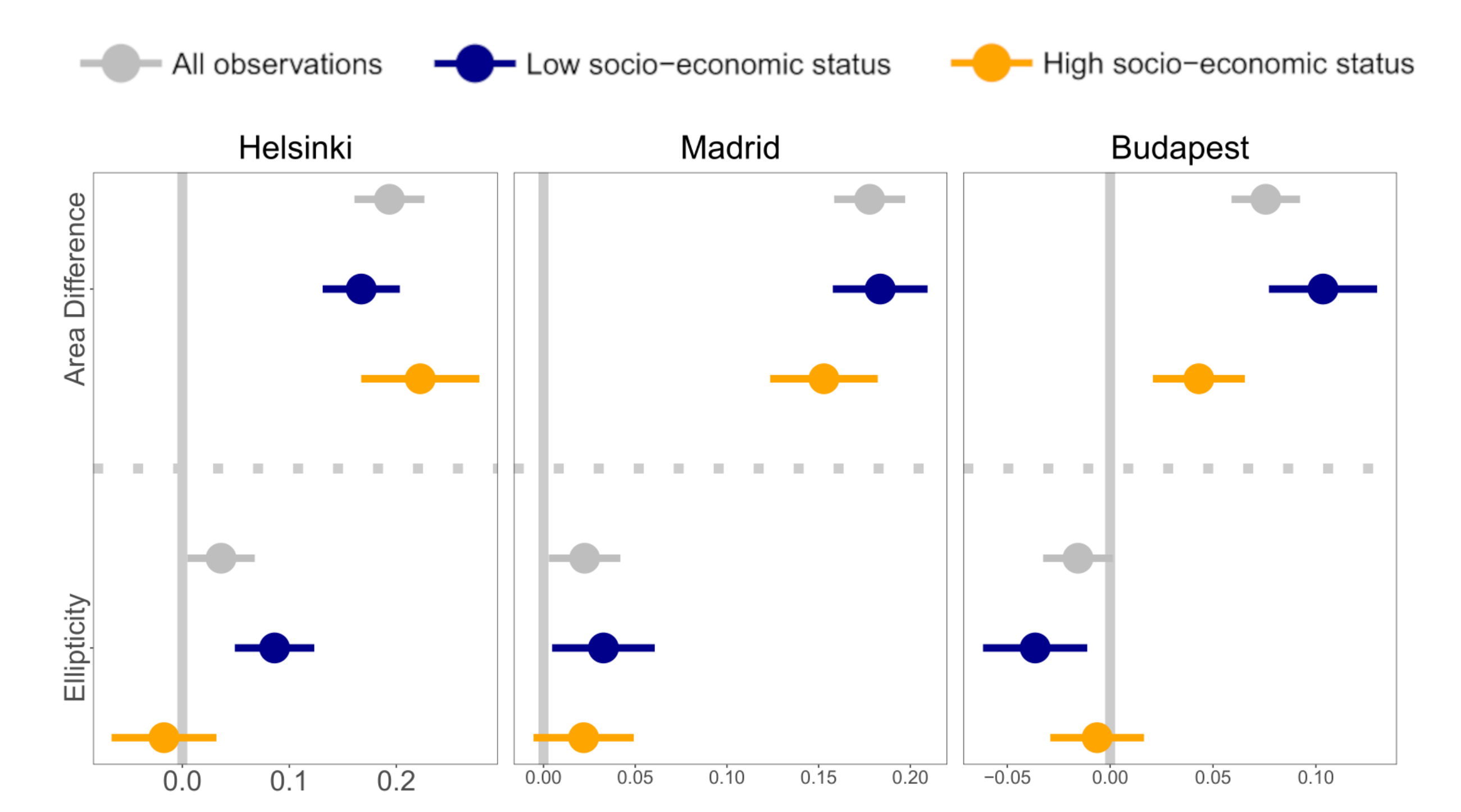}
        \caption{}
    \end{subfigure}

    \begin{subfigure}{0.25\linewidth}
        \includegraphics[width=\linewidth]{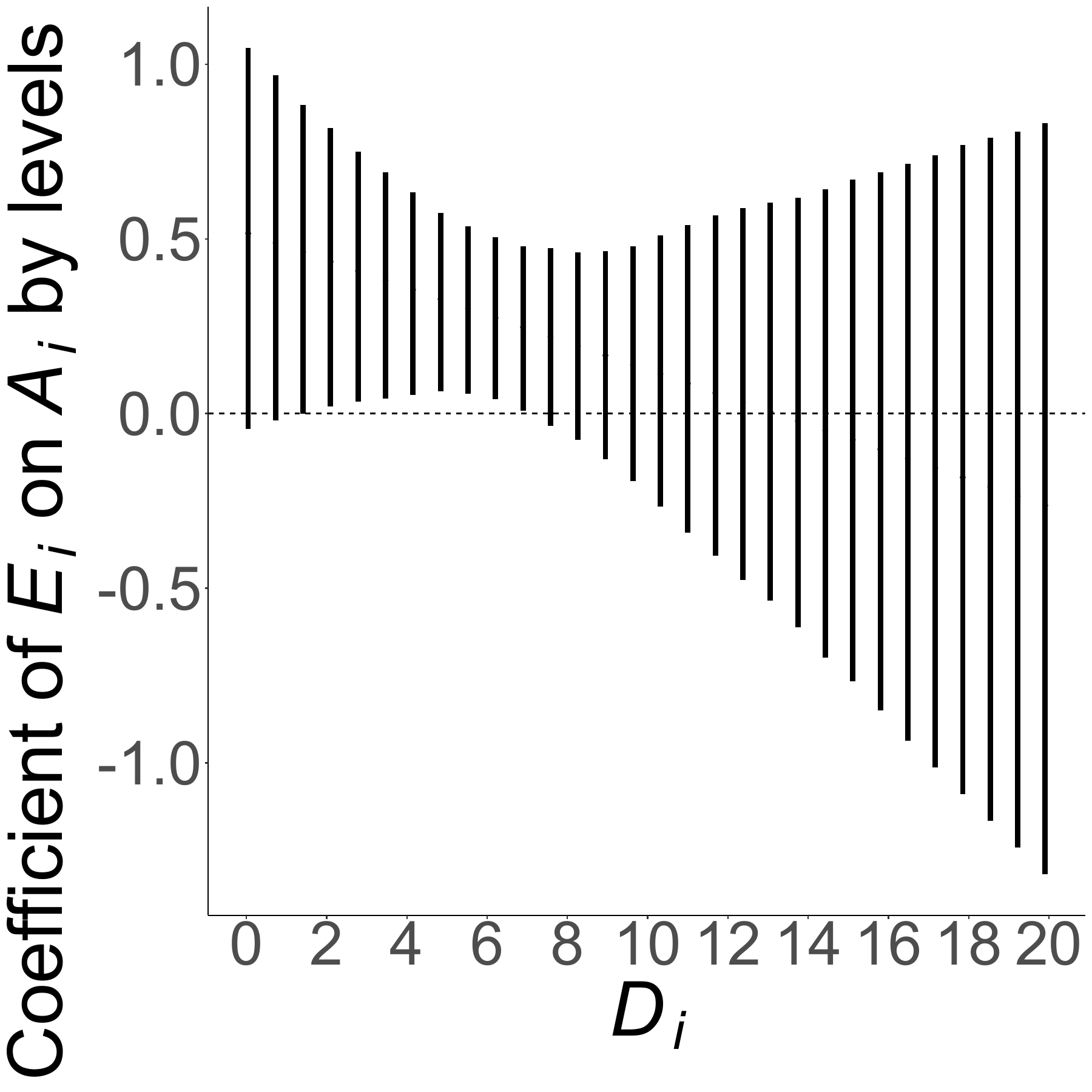}
        \caption{}
        \label{fig:interplot_helsinki}
    \end{subfigure}
    \begin{subfigure}{0.25\linewidth}
        \includegraphics[width=\linewidth]{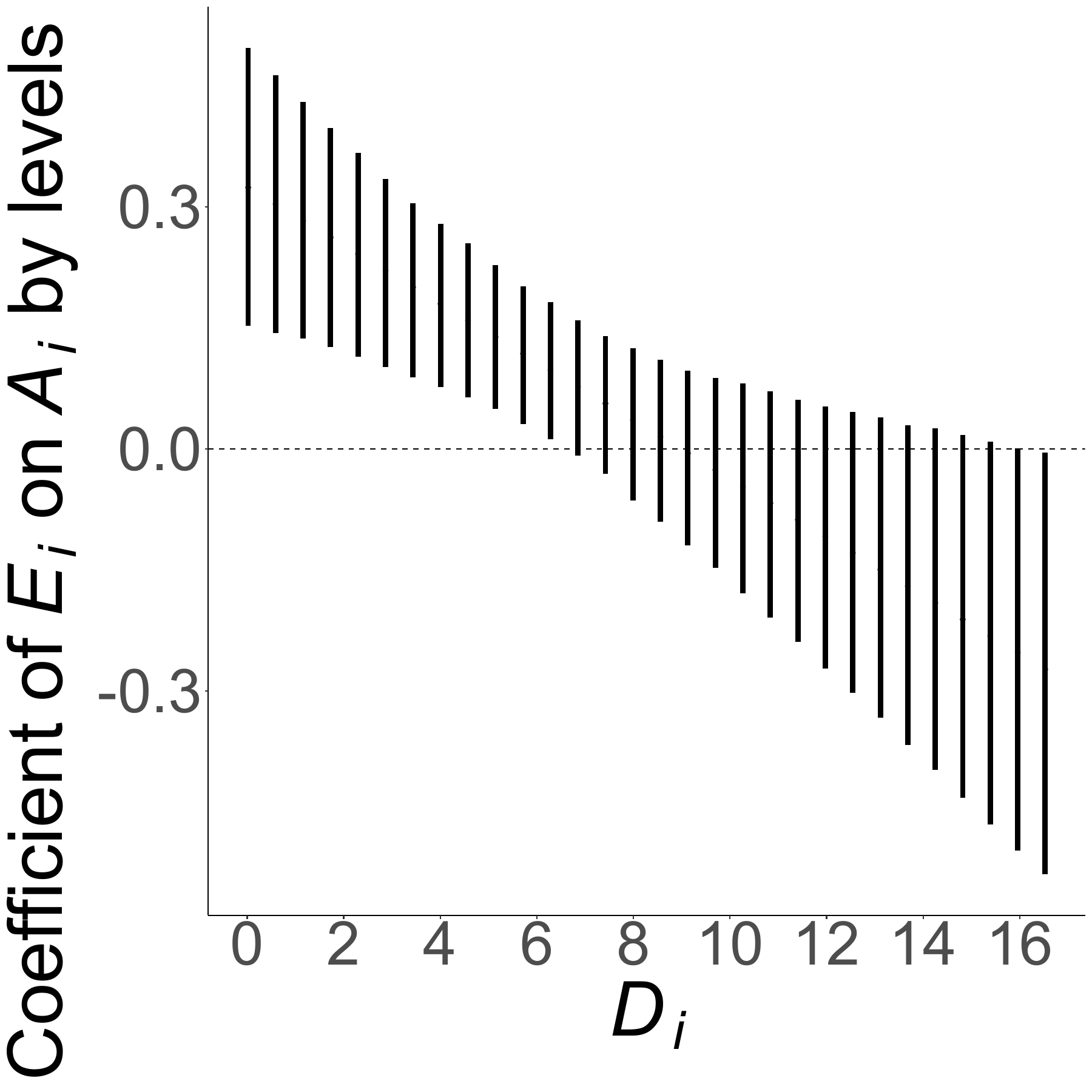}
        \caption{}
        \label{fig:interplot_madrid}
    \end{subfigure}
    \begin{subfigure}{0.25\linewidth}
        \includegraphics[width=\linewidth]{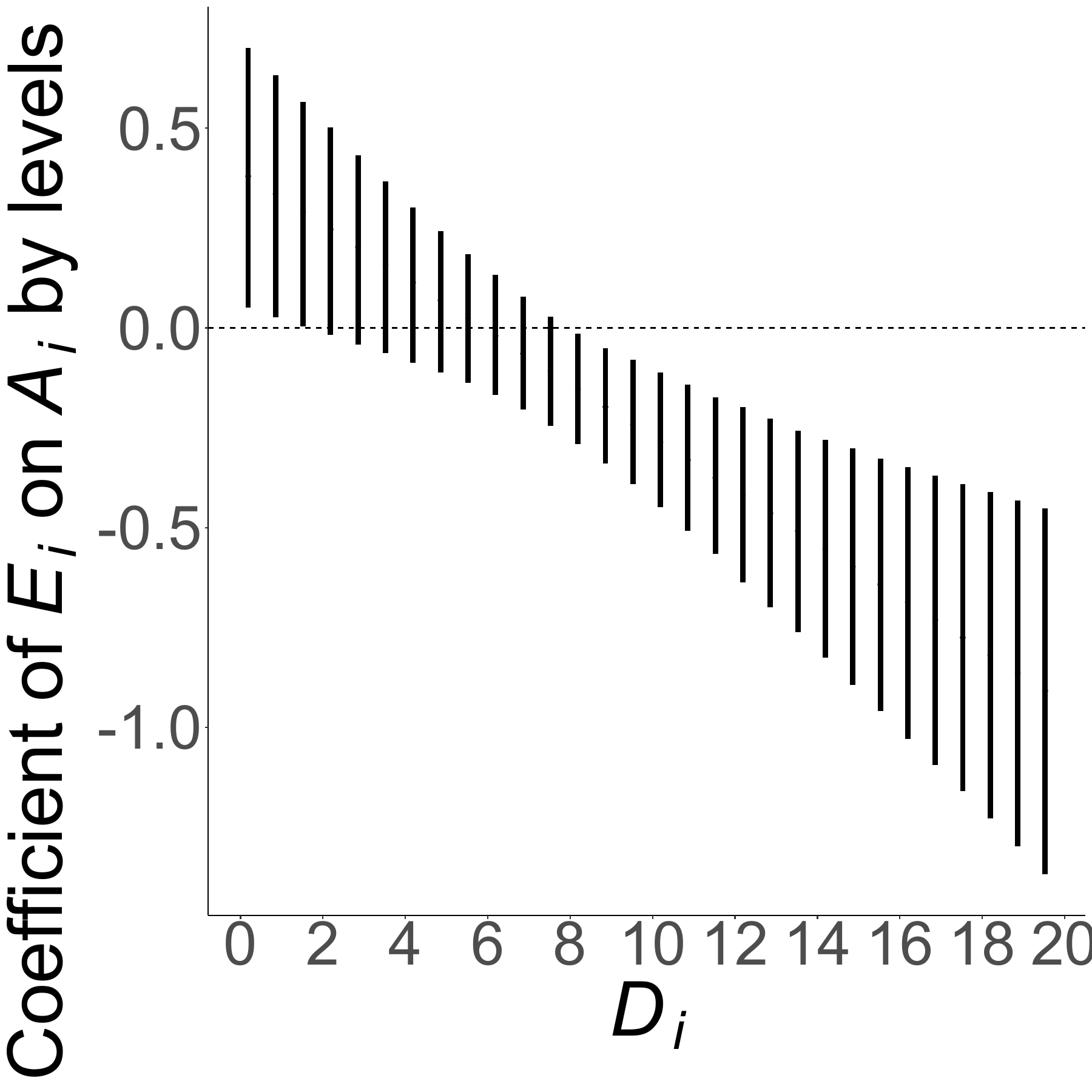}
        \caption{}
        \label{fig:interplot_budapest}
    \end{subfigure}
    \caption{\textbf{Amenity access in multimodal 15-minute cities.} \textbf{a)} We quantify the diversity of amenities that can be reached within a 15-minute walk to a multimodal radius. \textbf{b)} Multimodal reach increases amenity access. \textbf{c)} Regression coefficients, broken down to high- and low-status neighborhoods, highlight the role of multimodal mobility and elliptic transportation in increasing amenity access. Coefficients of ellipticity decreases in urban periphery in \textbf{d)} Helsinki, \textbf{e)} Madrid, and \textbf{f)} Budapest.}
    \label{fig:fig3}
\end{figure}

\begin{figure}[ht]
    \centering

    \begin{subfigure}{0.44\linewidth}
        \includegraphics[width=\linewidth]{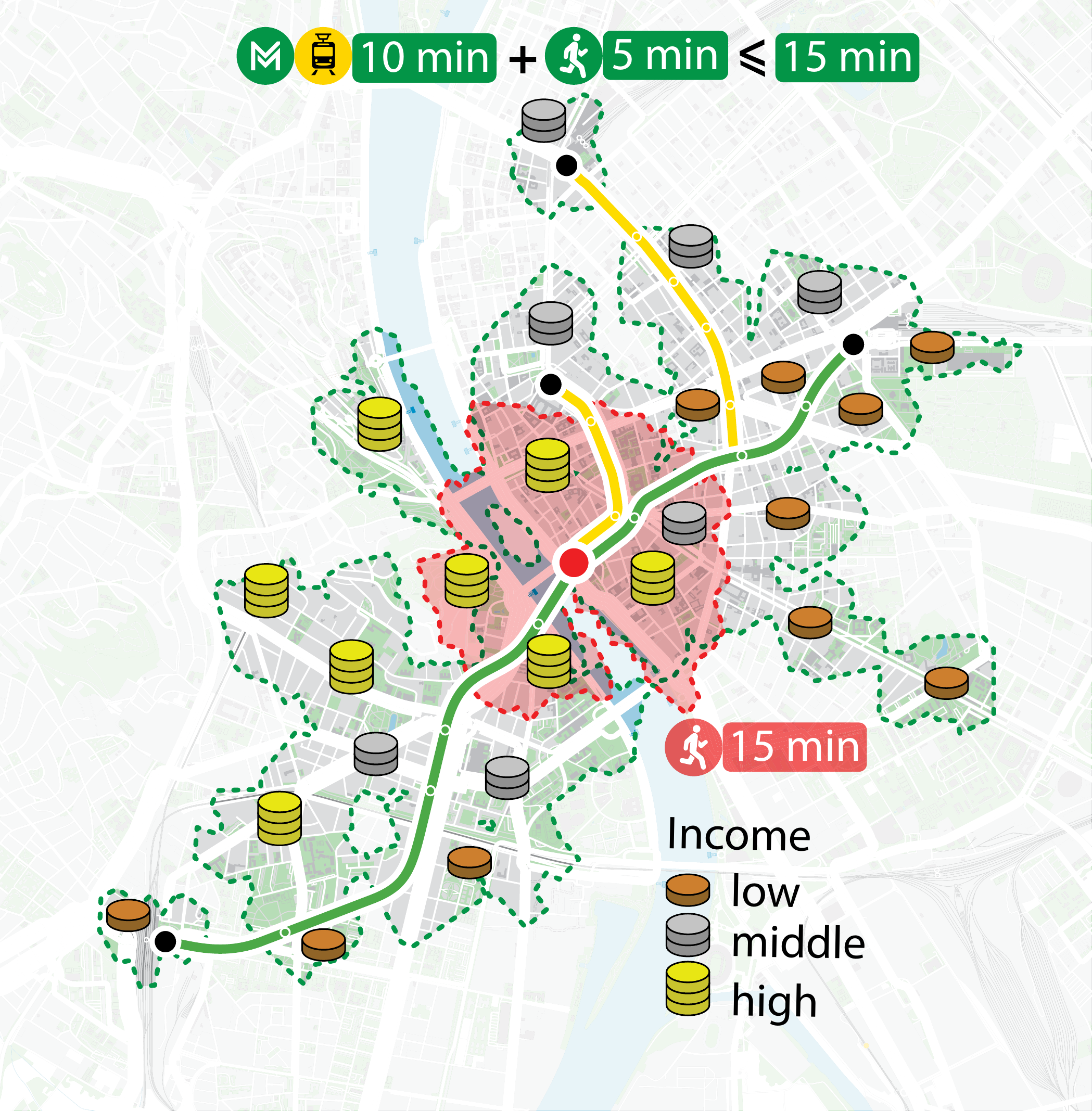}
        \caption{}
    \end{subfigure}
    \begin{subfigure}{0.44\linewidth}
        \includegraphics[width=\linewidth]{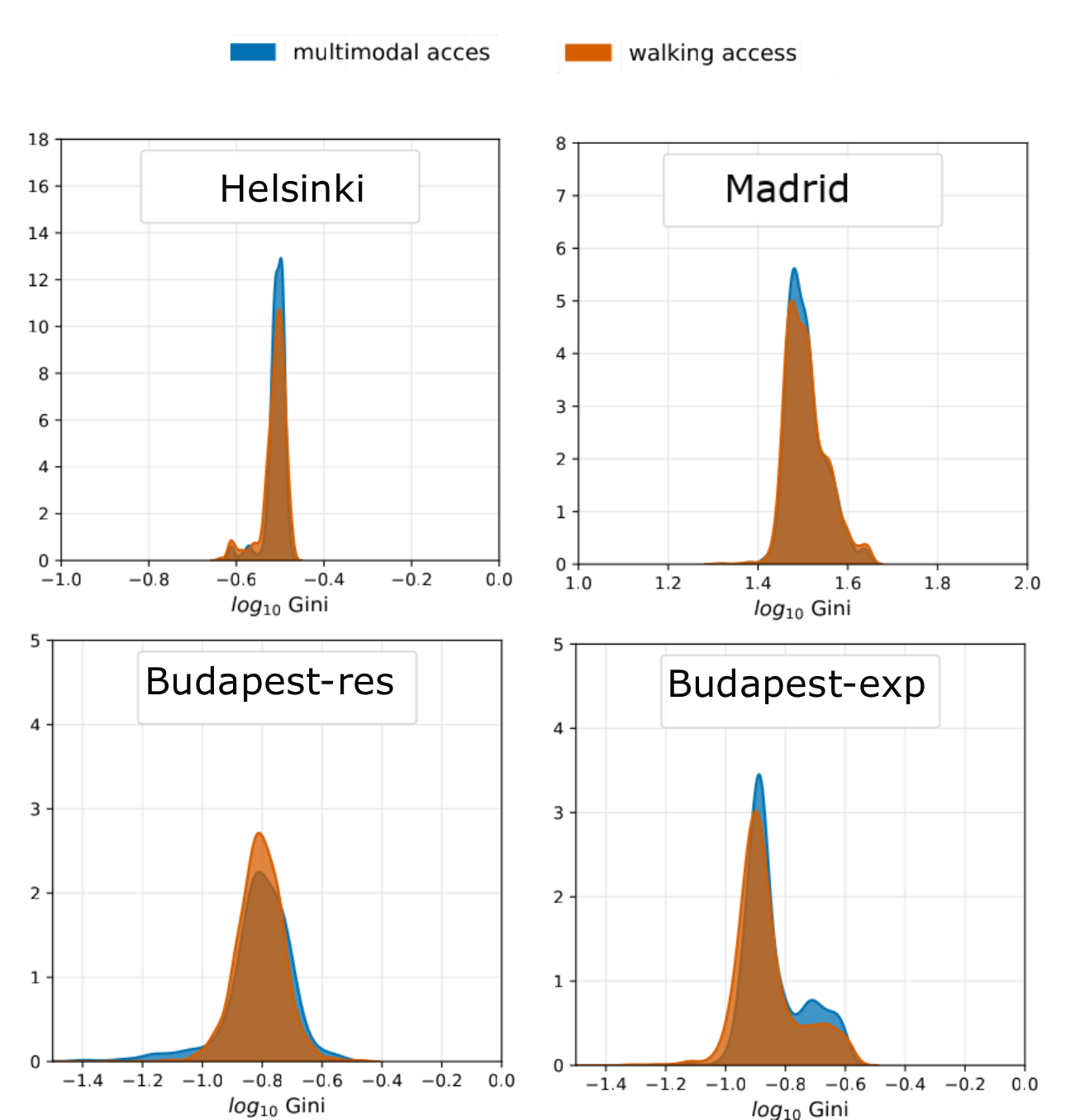}
        \caption{}
        \end{subfigure}

    \begin{subfigure}{0.72\linewidth}
        \includegraphics[width=\linewidth]{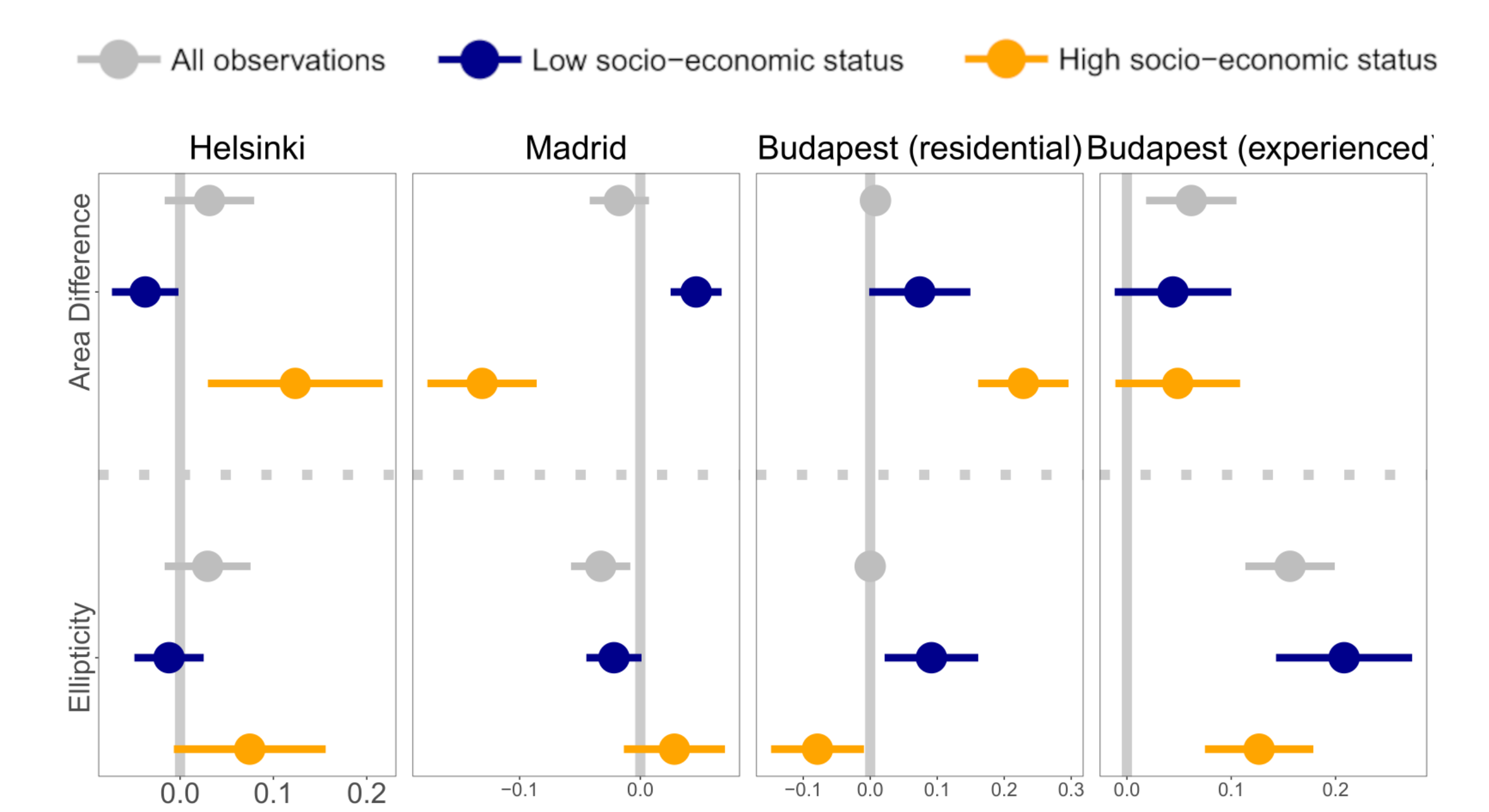}
        \caption{}
        \end{subfigure}

    \begin{subfigure}{0.21\linewidth}
        \includegraphics[width=\linewidth]{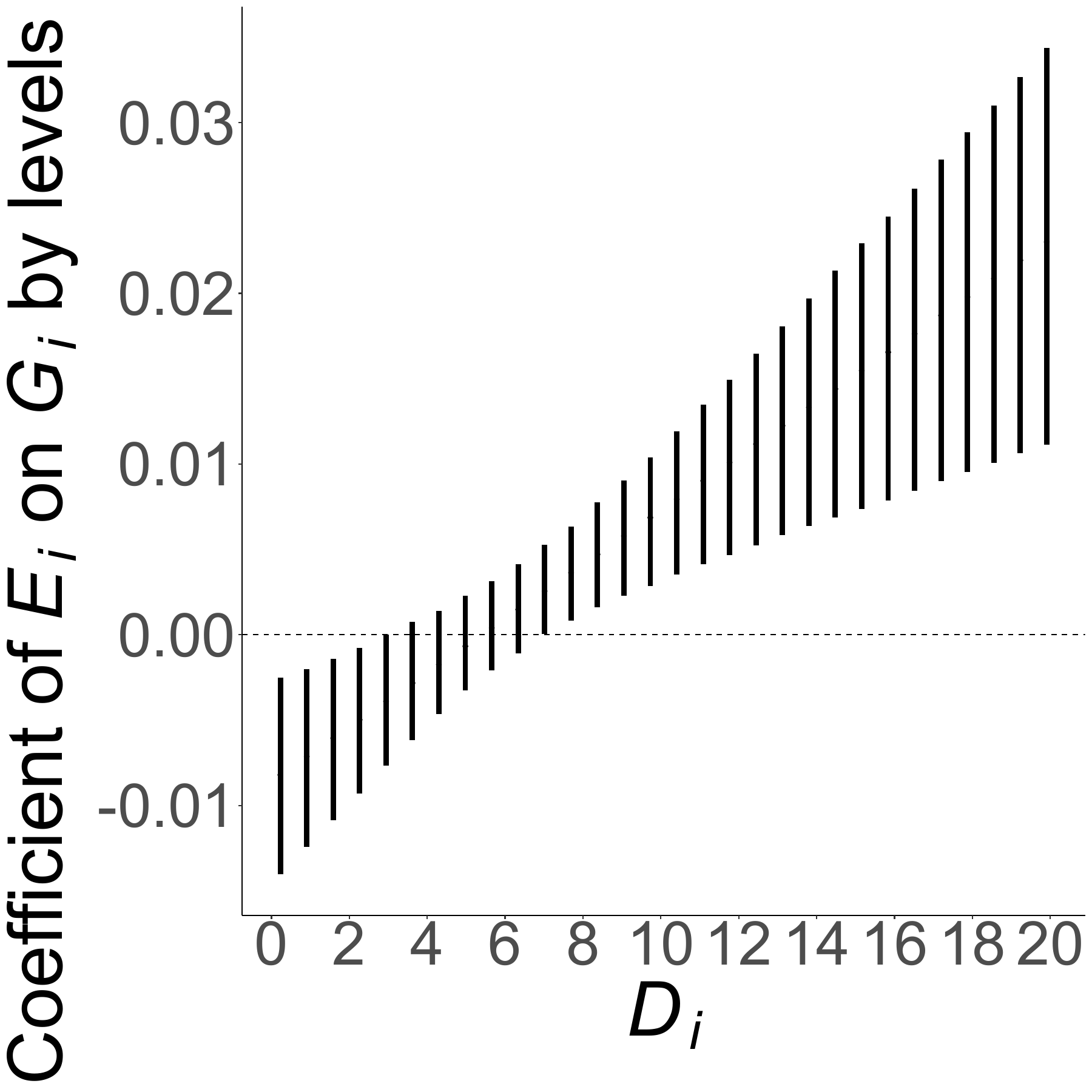}
        \caption{}
    \end{subfigure}
    \begin{subfigure}{0.21\linewidth}
        \includegraphics[width=\linewidth]{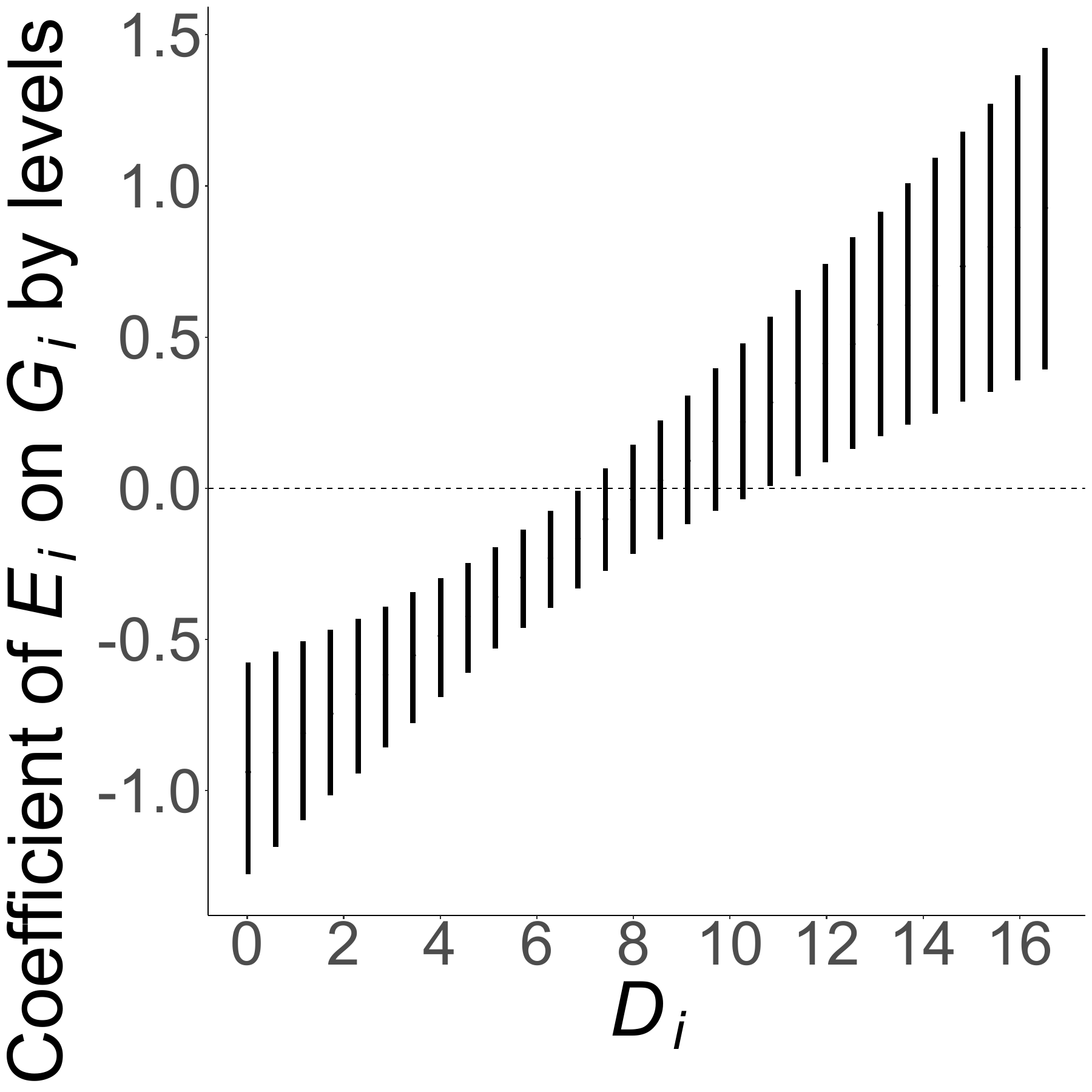}
        \caption{}
        \label{fig:gini_distribution_madrid}
    \end{subfigure}
    \begin{subfigure}{0.21\linewidth}
        \includegraphics[width=\linewidth]{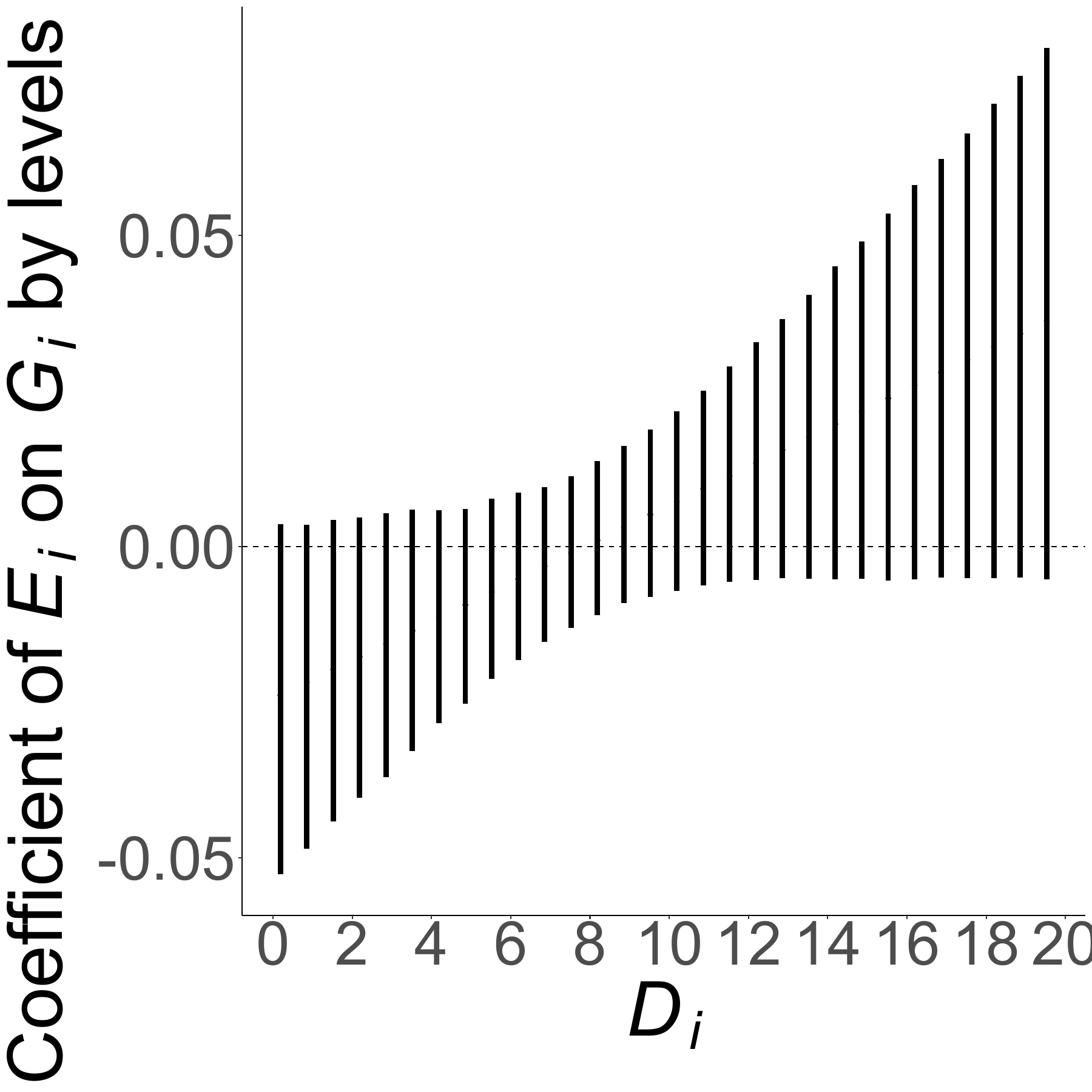}
        \caption{}
        \label{fig:gini_distribution_budapest}
    \end{subfigure}
    \begin{subfigure}{0.21\linewidth}
        \includegraphics[width=\linewidth]{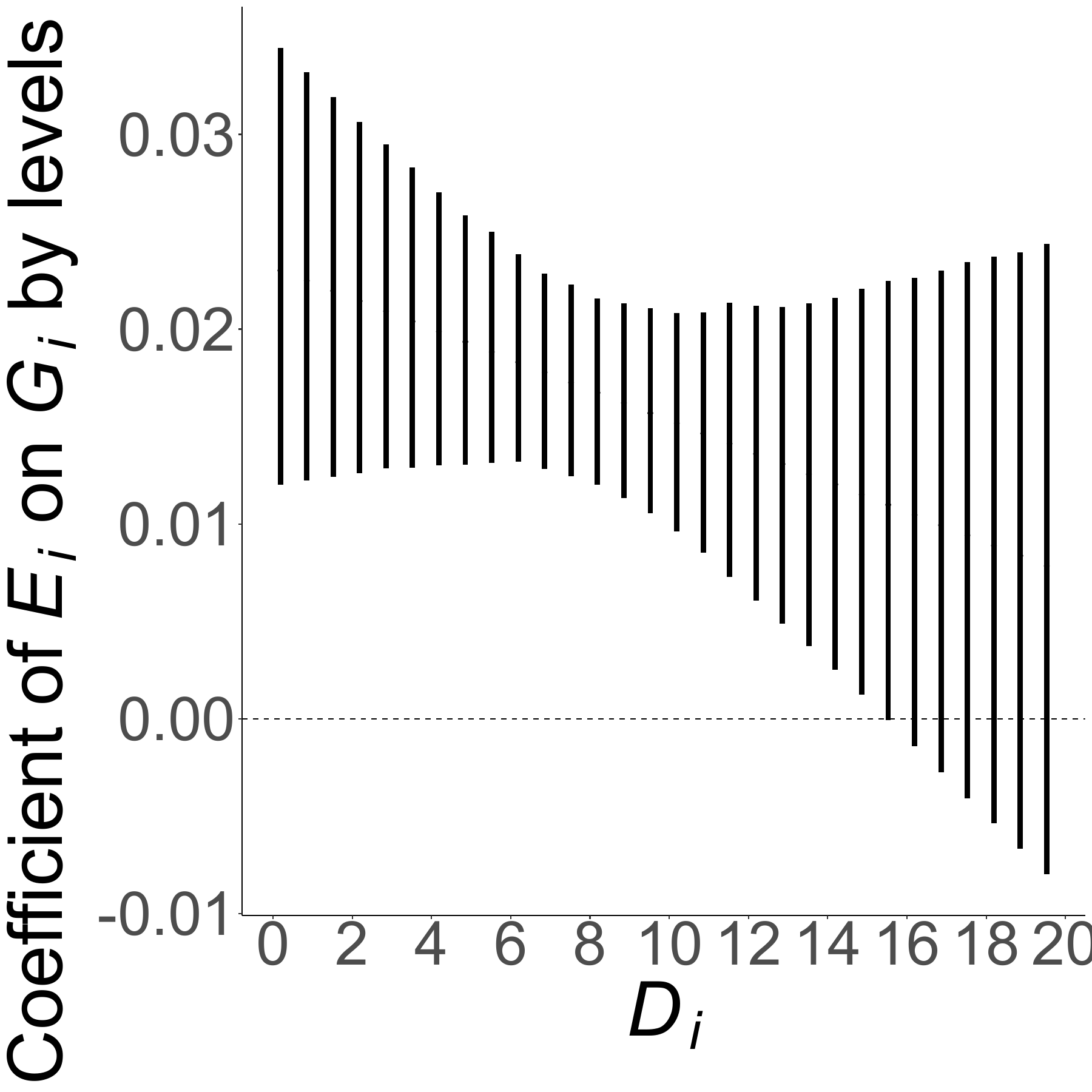}
        \caption{}
    \end{subfigure}

    \caption{\textbf{Social mixing potential in multimodal 15-minute cities.} \textbf{a)} We quantify the diversity of socio-economic status that can be reached within a 15-minute walk to a multimodal radius. \textbf{b)} Multimodal reach only slightly increases the mixing potential of socio-economic groups. \textbf{c)} Regression coefficients highlight that multimodal mobility and elliptic transportation increases social mixing potentials differently in high- and low-status neighborhoods. Coefficients of ellipticity increases in urban periphery for residential measures in \textbf{d)} Helsinki, \textbf{e)} Madrid, and \textbf{f)} Budapest, but not for experienced mixing in Budapest in \textbf{g)}.}
    \label{fig:fig_4}
\end{figure}

To quantify how public transport can improve amenity access and socio-economic mixing in 15-minute cities, we run a linear regression with the ordinary least squares method that is specified with the formula

\begin{equation}
    Y_{p\in(a,g)}=\alpha + \beta_1 S_p + \beta_2 E_p + \beta_3 D_p + \beta_5 X + \epsilon_p,
    \label{eq:eq1}
\end{equation}
where $Y_p$ correspond to our two dependent variables --amenity access and socio-economic mixing potential-- defined later in Figure \ref{fig:fig3} and Figure \ref{fig:fig_4}, $D_p$ is the Euclidean distance from the center of gravity in the city, $X$ is a vector of control variables including $Access_{walk}$ and $Mixing_{walk}$, values of amenity access and mixing potential in the walking area polygons (see Figure \ref{fig1}). Summary statistics and Pearson correlation values of the variables are reported in the Supplementary Information Section~\ref*{si:sec:descriptive_stats}. Despite the high correlation between $Y_p$ and controllers, we include variables of $X$ because their importance in understanding additional access provided by public transport. All variables are standardized by their means and standard deviations that allows comparison of coefficients. To examine whether the public transport premium varies by socio-economic context, we additionally run the regression separately for low- and high-income neighborhoods, defined by splitting the mobility hubs at the median average income of their 15-minute walking area (see Section~\ref{si:sec:socioeconomic_measure} of the Supplementary Information)

Next, we include the interaction term $E_p \times D_p$ to better understand how the coefficients of ellipticity depends on the distance from the city center: 

\begin{equation}
    Y_{p\in(a,g)}=\alpha + \beta_1 S_p + \beta_2 E_p + \beta_3 D_p + \beta_4 E_p \times D_p \beta_6 X + \epsilon_p.
    \label{eq:eq2}
\end{equation}
This will help understand how elliptic access improvement is related to amenity access and socio-economic mixing potential in areas close to city center and in peripheral neighborhoods. Then, we can calculate the marginal effect of ellipticity based on the distance from the center $    \frac{\partial Y_p}{\partial E_p} = \beta_2 + \beta_4 D_p$. This marginal effect indicates that the contribution of ellipticity varies systematically with distance from the city center. A positive $\beta_4$ implies that the effect of ellipticity becomes stronger in more peripheral areas, whereas a negative $\beta_4$ suggests that elliptic accessibility matters more in central urban locations. This technique also enables us to quantify how the magnitude and significance of relationship between ellipticity and the dependent variable changes along the distance from the city center.

In the first analysis, we investigate how many new types of amenities are accessible within 15-minutes if we allow 10 minutes of public transport travel compared to the 15-minute walk accessibility. The dependent variable is $Y_{p,a}=Access_{multimodal}-Access_{walk}$, where $Access$ quantifies the number of available amenity types (Figure \ref{fig:fig3}a) that are considered essential for local access in the 15-minute city literature \cite{abbiasov202415} (see Section~\ref{si:sec:amenity_access_from_osm} of the Supplementary Information). As the number of available amenities are considerably higher in the multimodal than in the walking accessibility polygons (Figure \ref{fig:fig3}b), we can expect increase in available amenity types too.

Indeed, the regression analysis following Equation \ref{eq:eq1} confirms that multimodal 15-minute polygons provide access to significantly more amenity types than walking-only catchments across all three cities  (Figure \ref{fig:fig3}c). The larger $S_p$ the greater the amenity gain that signals an importance of efficient public transport that enables citizens to cover larger areas in their 15-minute multimodal trip. The decomposition of the regressions to high- and low socio-economic status neighborhoods -- based on their income or real estate prices, see Materials and Methods for details -- confirms that this additional access is distributed equally across income groups in Madrid and Helsinki, while the lower status group benefits more from multimodal access in Budapest.

To understand whether the benefit of elongated multimodal access is equally distributed across income groups, we examine the relationship between $E_p$ and $Y_{p,a}$ (Figure \ref{fig:fig3}c). In Madrid and Helsinki, elliptical multimodal polygons increase access to low-status neighborhoods without significantly affecting high-status ones. This suggests that unidirectional transit can partially compensate for deficits in walkability in disadvantaged areas. In Budapest, however, ellipticity is associated with fewer new amenities in disadvantaged areas compared to round-shaped polygons, indicating that the equity benefits of unidirectional public transport are not universal and depend critically on the spatial structure of the local transport network.


Finally, we investigate the relationship between the multimodal polygon shape $E_p$ and amenity access $Y_{p,a}$ as a function of distance from the center $D_p$ using Equation \ref{eq:eq2}. We find that the coefficient of $E_p$ on amenity access decreases monotonically with $D_p$ across all analyzed cities (Figure \ref{fig:fig3}d-e-f). In peripheral areas of Helsinki and Madrid, the shape of multimodal reach does not change the accessible amenities, contrary to the expectation that peripheral areas would benefit from fast radial access towards the center. Furthermore, in peripheral Budapest, ellipticity has a significantly negative relationship with amenity access.
These findings highlight the importance of polycentric public transport in urban peripheries to increase accessibility to amenities within a short reach. We come back to this point in the Discussion section.

Although improving access to amenities is a central goal of the 15-minute city, a common criticism is that it may reinforce existing socio-economic segregation. To examine whether public transport expands exposure to socio-economically diverse neighborhoods, we compare socio-economic mixing potential in 15-minute walking with multimodal mobility polygons (Figure \ref{fig:fig_4}a). We use the Gini index to measure the socio-economic diversity of neighborhoods reachable within each polygon, where 
higher values of $G_p$ indicate that a given location can reach a more socio-economically diverse set of neighborhoods, making it a positive outcome in this framework.

The distributions of the Gini index reveal that multimodal mobility only moderately changes the socio-economic composition of accessible neighborhoods compared to 15-minute walking polygons (Figure \ref{fig:fig_4}c). In all three cities, multimodal access shifts the distribution of reachable socio-economic diversity only slightly relative to walking access, indicating that public transport expands access mostly within already similar socio-economic environments. Nevertheless, the shift is consistently visible toward higher Gini values. The effect is most pronounced in Budapest when considering residential rather than experienced mixing, suggesting that public transport creates meaningful, albeit modest, opportunities to access socially diverse areas within a short travel radius.

Whether multimodal mobility promotes social mixing depends strongly on the socio-spatial organization of each city. 
The regression coefficients of the additional multimodal area $S_p$ reverse sign between low- and high-status neighborhoods in Helsinki and Madrid, indicating opposing effects on social mixing potential(Figure \ref{fig:fig_4}c). In Budapest, however, multimodal mobility increases mixing potential regardless of neighborhood status. The role of ellipticity $E_p$ is similar across urban contexts. In Helsinki and Madrid, elliptic multimodal access is positively associated with mixing in high-status neighborhoods, while the effect is absent in low-status areas. In Budapest, the effect is stronger in lower-status locations for both residential and experienced measures. These differences suggest that rapid corridor-type public transport enhance social mixing where it connects socially differentiated urban zones, but its effectiveness depends strongly on the socio-spatial organization of each city.

Radial public transport corridors become increasingly effective for residential social mixing in peripheral neighborhoods, revealing a spatial gradient absent from the amenity access results (Figure \ref{fig:fig_4}d-e-f). Across all three cities, when looking at the residential data, the coefficient of ellipticity becomes increasingly positive in peripheral neighborhoods, indicating that unidirectional public transport corridors are more effective for social mixing in outer urban areas than near the center. For Helsinki (Figure \ref{fig:fig_4}d) and Madrid (Figure \ref{fig:fig_4}e), this relation is statistically significant; Budapest displays a weaker but still positive peripheral tendency (Figure \ref{fig:fig_4}f). This pattern contrasts with amenity access results, where ellipticity loses effectiveness toward the periphery. However, experienced mixing in Budapest tells a different story: the marginal effect of ellipticity declines with distance (Figure \ref{fig:fig_4}g), implying that although radial public transport corridors connect peripheral residential areas to socially different places, they do not necessarily translate into experienced encounters in everyday mobility. Together, these findings suggest that fast radial transport lines can support socio-economic integration in peripheral neighborhoods even when they do not substantially improve access to diverse amenities.

\clearpage

\section{Discussion}
\label{sec:discussion}

Our analysis demonstrates that incorporating public transport into the 15-minute city framework substantially changes how local accessibility should be understood. Across Helsinki, Madrid, and Budapest, multimodal mobility consistently expanded access to amenity types within a short temporal reach, with the strongest gains observed where public transport enlarged the accessible area most efficiently. At the same time, improvements in socio-economic mixing were more moderate and depended strongly on the socio-spatial structure of each city. While radial or corridor-like transport lines often increased the probability of reaching socially different neighborhoods, especially in peripheral areas, their contribution to amenity access was weaker and in some cases negative. These findings indicate that the spatial geometry of public transport matters beyond simple travel time reduction: the same transport extension can generate very different social and functional accessibility outcomes depending on whether it broadens access in multiple directions or primarily along a single axis.

These results speak directly to one of the main unresolved questions of the 15-minute city debate: how proximity-based planning can remain socially inclusive outside dense urban cores. The original concept assumes that most daily needs should be reachable locally, yet urban peripheries often lack sufficient amenity diversity and rely structurally on public transport. Our findings suggest that public transport should not be treated merely as a compensatory layer added to walking accessibility, but as a structural component of short-reach urban mobility. In particular, the limited amenity gains generated by elliptic, predominantly radial transit corridors highlight the need for polycentric transport systems that connect peripheral neighborhoods not only to central business districts but also horizontally to secondary urban centers. Such networks may better complement active mobility by distributing access opportunities across multiple destinations rather than reinforcing monocentric movement patterns. Helsinki is a good example of this: its coastal geography forces radial lines into elliptic shapes that cannot expand laterally. This constrains access to amenities regardless of service quality. Geographic context shapes what transport networks can deliver.

Several limitations should be considered when interpreting these results. First, the analysis relies on scheduled GTFS travel times and therefore cannot capture congestion, service unreliability, or temporal variation in waiting times that may substantially affect realized accessibility. Additionally, the model assumes boarding upon arrival at a stop, without accounting for timetables, which might reduce the realized access premium compared to our estimates. Second, socio-economic mixing is approximated through residential income, real estate prices, and in Budapest partly through temporally observed population composition, which represent different dimensions of social exposure and are not fully comparable across cities. Third, the 15-minute threshold remains a normative simplification, while actual travel tolerances vary across population groups and trip purposes. Future research could extend this framework by integrating observed mobility traces, dynamic service frequencies, the hierarchical layers of amenities, the complementary role of public transport and shared mobility, and additional cities with different transport morphologies. A particularly important next step is to investigate whether multimodal accessibility not only increases potential encounters but also translates into repeated social interaction and long-term reductions in urban segregation.

\section{Materials and Methods}
\label{sec:materials_and_methods}

\subsection{Data sources}
\label{sec:data_sources}

The urban data used to create the public transport network, the walking areas and the amenity categories is publicly available. We use \acrfull{GTFS}, an open-standard format used by public transit agencies globally to publish scheduled data, including routes, stops and timetables. Walking isochrones are computed from street network data sourced from Geofabrik. \acrfull{POI} are extracted from \acrfull{OSM}, a global geographic database contributed to by volunteers, and aggregated into higher-level amenity categories following \cite{abbiasov202415}. (see SI Section S2).

Socio-economic data are sourced at the neighborhood level and vary by city according to data availability. For Helsinki, we use Paavo open data published by Statistics Finland at the postal code level. For Madrid, income and Gini coefficients are available directly at the census tract level from the Spanish National Statistics Institute (INE), accessed via ineAtlas. Both sources are freely available and provide the income and inequality estimates used to construct the social mixing variable and to define income subgroups in the regression.

Hungary does not publish comparable open income data at fine spatial scales. 
For the residential mixing measure, we use property listing data from ingatlan.com, a Hungarian real estate platform comprising approximately 62,000 listings within Budapest from 2018. The dataset was provided to our research group by the company in 2018. From these listings, we derive square meter prices as a proxy for neighborhood socioeconomic status. For the experienced mixing measure, we use \acrfull{TPDD} from Magyar Telekom, a leading Hungarian telecommunications service provider. This dataset is accessible through an data purchase contract and an NDA between Magyar Telekom and ELTE KRTK. The data is anonymous, its aggregation does not allow identification of indivduals but provides hourly, aggregated population counts by income group (low, middle, and high) at a 100$\times$100\,m hexagonal grid, derived from mobile network activity.

\subsection{Network construction and calculating access area}
We construct city-specific public transportation networks using \acrfull{GTFS} using the most densely scheduled hour on a representative weekday for each city. We cluster spatially proximate stops into mobility hubs to enable realistic transfers (see Section~\ref*{si:sec:creating_public_transport_network} of the Supplementary Information). The resulting network is modeled as a directed multigraph, where nodes correspond to mobility hubs and weighted edges represent scheduled travel times. We apply a transfer penalty of 180 seconds when switching between vehicles. We identify all reachable stops within a 10-minute travel budget from each node using a depth-first search. From each node, we use a depth-first search to identify all reachable stops within a 10-minute travel budget, producing a node-to-node reachability list that forms the basis of the multimodal access area calculation.

Access areas are computed for both walking and multimodal mobility using Valhalla \cite{valhalla320} as the routing engine. For the walking condition, we calculate 15-minute walking isochrones directly from each stop. For the multimodal condition, we first identify all stops reachable within ten minutes by public transportation. Then, we compute five-minute walking isochrones from each of those stops and take their union as the multimodal catchment area. The difference between the multimodal and walking catchments, in terms of covered area and socio-economic composition of reachable neighborhoods, forms the basis of our comparative analysis.

\subsection{Metrics}
\label{sec:metrics}

To measure socio-economic mixing within each catchment, we compute the Gini 
coefficient
\begin{equation}
G_p = \frac{\sum_{i=1}^{n}\sum_{j=1}^{n} |x_i - x_j|}{2n^2 \mu}
\end{equation}
where $x_i$ denotes the income proxy of neighborhood $i$, $n$ is the number of neighborhoods within polygon $p$, and $\mu$ is their mean. The income proxy varies by city and data availability (see Section~\ref*{si:sec:socioeconomic_measure} of the supplementary Information). Gini coefficients are aggregated to the catchment polygon level using area-weighted averaging across intersecting administrative units. Our key outcome variable, $\Delta G_p$, is the difference between the Gini coefficient of the multimodal catchment and that of the 15-minute walking catchment, capturing the change in socio-economic mixing attributable to public transport access.


To characterize the shape of the multimodal access area, we define the ellipticity of each catchment \( E \). First, the covariance matrix \( C \) of the points (the geographic coordinates of the accessible stops) is computed (\ref{eq:covariance_matrix}).

\begin{equation}
    \label{eq:covariance_matrix}
   C = \begin{pmatrix}
   \text{Cov}(x, x) & \text{Cov}(x, y) \\
   \text{Cov}(y, x) & \text{Cov}(y, y)
   \end{pmatrix}
\end{equation}

Second, the eigenvalues \( \lambda_1 \) and \( \lambda_2 \) of the covariance matrix \( C \) are calculated.

Third, the lengths of the semi-major and semi-minor axes are given by $a = \sqrt{\lambda_1}$ and $ \quad b = \sqrt{\lambda_2}$.
Finally, the ellipticity \( E \) is computed as $E = 1 - \frac{b}{a}$, or substituting $a$, and $b$: (\ref{eq:ellipticity}).

\begin{equation}
\label{eq:ellipticity}
E = 1 - \frac{\sqrt{\lambda_2}}{\sqrt{\lambda_1}}
\end{equation}

The ellipticity is computed for every stop cluster.
$E_p$ ranges from 0 (perfectly circular coverage) to 1 (maximally elongated), with higher values indicating that the PT network extends predominantly in one direction from the origin stop.
Figure~\ref{fig:ellipticity_examples} shows two examples, a rather circular one downtown (''Margit híd, budai hídfő'') with the ellipticity of \num{0.1142}.
The example is an elongated access area with the ellipticity of \num{0.9134} from the outskirts (''513. utca'').

To locate each stop relative to the urban core, we identify the city center 
using betweenness centrality. We compute the Euclidean distance in kilometers from each stop to the mobility hub with the highest centrality, which yields the variable $D_i$, which we use in the regression analysis to test whether the effect of the integration of public transport on socio-economic mixing varies with distance from the center.

\subsection{Regression framework}
\label{sec:regression_framework}

We estimate the two regression models specified in Equations~\ref{eq:eq1} 
and~\ref{eq:eq2} separately for amenity access ($Y_{p,a}$) and socio-economic 
mixing ($Y_{p,g}$), and for each city. The dependent variables are constructed 
as the difference between the multimodal and walking area outcomes: 
$Y_{p,a} = Access_{multimodal} - Access_{walk}$ counts the gain in available 
amenity types, and $Y_{p,g} = G_{multimodal} - G_{walk}$ is the change in the 
Gini coefficient (see Metrics). All variables are standardized by their Z-scores.

To examine whether the public transport premium varies by socio-economic 
context, we additionally run each regression separately for low- and 
high-income subsamples. Stops are assigned to income groups based on the 
average socio-economic status of their 15-minute walking polygon. We split at 
the median income variable of the walking polygon (see variable definitions 
for each city in SI Section~\ref*{si:sec:socioeconomic_measure}).

To quantify how the contribution of ellipticity varies with distance from the 
city center, we evaluate the marginal effect $\frac{\partial Y_p}{\partial E_p} 
= \beta_2 + \beta_4 D_p$ from Equation~\ref{eq:eq2} based on distance from the center  in each city. At successive distance levels, we report the estimated marginal effect together with its 95\% confidence interval.

\section*{Declarations}


\subsection*{Acknowledgments}
The authors wish to thank the ANETI Lab Brown Bag seminar for their comments and suggestions. We acknowledge the assistance of our illustrator Szabolcs Tóth-Zs. in finalizing our primary figures. 
Z.Z. and G.P. acknowledge funding from the Bridge grant (2024) from yrCSS. Z.Z. thanks the ANETI lab for hosting her during part of the research. B.L. acknowledges financial help received from the MTA Lendület Award. The authors acknowledge financial help from NRDI Office of Hungary through the Driving Urban Transition European Partnership project COLINE (Complex Links of Neighbourhoods).

\subsection*{Ethics declarations}
Competing interests

The authors declare no competing interests.

\subsection*{Data availability}

We combine multiple data in this study, and most of them are openly available.

Map data is from OpenStreetMap, copyrighted by the OpenStreetMap contributors and
licensed under the Open Data Commons Open Database License (ODbL).
The GTFS data used to construct the transportation networks are also openly available via the public transport companies of Budapest, Helsinki, and Madrid.

The socio-economic data used for Finland and Spain are open data, and available under CC-BY-4.0.
The Spanish income data is from INE, Instituto Nacional de Estadística via ineAtlas.data, and the Finnish data is obtained from the Paavo database by Statistics Finland.

The temporal population density data for Budapest was obtained from Magyar Telekom, and the property price data were provided by the ingatlan.com estate selling portal.
These datasets cannot be shared publicly.

\subsection*{Code availability}

The code developed for the study is available on GitHub in two repositories: i) the GTFS processing by M.M. in \url{https://github.com/mmate98/gtfs_multilin_analyses}, and ii) the rest of the workflow in \url{https://github.com/ANETILab/public_transport_in_the_15minute_city}.

\subsection*{Author contribution}
Z.Z., G.P., and B.L. designed the analysis. B.K. and I.F. prepared the data; Z.Z., G.P., M.M., and B.L. performed the analysis. Coding: G.P. (OSM processing and research software engineering), M.M. (GTFS processing), Z.Z. (Gini calculations and regressions), and B.L. (regressions). Z.Z and B.L wrote the paper; G.P. and M.M. contributed to the Supplementary Information. B.L supervised the project. All authors discussed the results and contributed to the final manuscript.

\printbibliography[title=References]
\end{refsection}

\clearpage
\section*{Supplementary Materials}

\renewcommand{\thefigure}{S\arabic{figure}}
\renewcommand{\thetable}{S\arabic{table}}
\renewcommand{\thesection}{S\arabic{section}}
\renewcommand{\theequation}{S\arabic{equation}}
\setcounter{figure}{0}
\setcounter{table}{0}
\setcounter{section}{0}
\setcounter{equation}{0}

\begin{refsection}

\section{Creation of the public transport network}
\label{si:sec:creating_public_transport_network}

\subsection{Clustering}

We performed clustering of public transport stops for two main reasons. First, clustering allowed us to evaluate the significance of stop groups or hubs within each network, providing insight into the importance of specific locations in overall network connectivity, by calculating their centrality values.

Second, stop clustering was essential to ensure a fully connected network structure. When analyzing accessibility from a given node, it is important to allow the algorithm to recognize when transfers between nearby stops are easy and realistic. By grouping spatially proximate stops into clusters, we could introduce artificial transfer edges between them, effectively simulating walking connections. We use a different approach for Budapest, where we have local knowledge on the transportation system and can qualitatively assess the validity of the clusters, and Madrid and Helsinki, where we rely on an optimization approach to ensure validity.

\subsubsection{Budapest}

In the case of Budapest, each stop's coordinates are first converted into a geodataframe, and then a full pairwise geodesic distance matrix $D$ is computed, in which $D_{ij}$ correspond to the geodesic distance between stops $i$ and $j$.

This matrix measures the real-world surface distances (in meters) between all stops using latitude and longitude coordinates. Based on this distance matrix, we apply agglomerative hierarchical clustering with a 150 meters distance threshold, aligning with \cite{kocsis2022extracting} observation that stop counts drop sharply between 100–200 meters and supported by our own finding that this distance best captures functionally close stations. Stops are clustered so that no two within the same cluster are more than 150 meters apart. Isolated stops form single-stop clusters, while others are grouped with nearby stops.
While the 150 meters threshold generally creates reasonable clusters, we applied a manual post-processing step to merge stops with the same name. This is particularly important at hubs with multiple intersecting lines, where spatial separation may slightly exceed the threshold (typically in the case of large interchange stations). This adjustment ensures that these key transfer points are clustered appropriately, reflecting their role in network connectivity.

\subsubsection{Helsinki and Madrid}

For Helsinki and Madrid, a different approach was required due to the group's limited local knowledge in adjusting clusters and naming conventions. Here, clustering was based both on spatial distance and on the similarity of stop names. For Helsinki, we also removed certain suffixes representing transport types (e.g., “M”) to facilitate more accurate name embedding.
We used the relevant spaCy language models (Finnish and Spanish, respectively) to embed stop names and compute cosine similarity between their vector representations.
Both pairwise distance matrix and the name similarity matrix were normalized to a common scale. The final similarity metric for clustering was computed as follows \cite{hammer2025bayesian}:

\begin{equation}
 S_{ij}=\alpha \times distance^{norm}_{ij}+(1-\alpha)\times (1-namesim_{ij})
\end{equation}

where $S_{ij}$ is the combined score between $stop_i$ and $stop_j$, $\alpha$ is a weight parameter controlling the importance of spatial distance and name similarity, $distance^{norm}_{ij}$ is the normalized distance between $stop_i$ and $stop_j$, $namesim_{ij}$ is the normalized cosine similarity between name embeddings.

Since cosine similarity is higher for more similar names, we used $1-namesim$ to ensure that higher values correspond to greater dissimilarity.
After combining the matrices, we explored different values of alpha and clustering thresholds. To select the optimal agglomerative hierarchical clustering configuration with complete linking, we systematically evaluated the silhouette score for each parameter combination on the original distance matrix. The silhouette score measures how similar each stop is to its own cluster compared to other clusters, providing a quantitative assessment of cluster quality.
To further refine our parameter selection and account for model uncertainty, we applied a Bayesian optimization approach over the parameter space ($\alpha$ and threshold). This allowed us to efficiently identify the clustering configuration with the highest expected silhouette score, balancing spatial proximity and semantic similarity in a principled, data-driven way (Hammer et al., 2025).

\begin{figure}[t!]
\centering
\includegraphics[width=\textwidth]{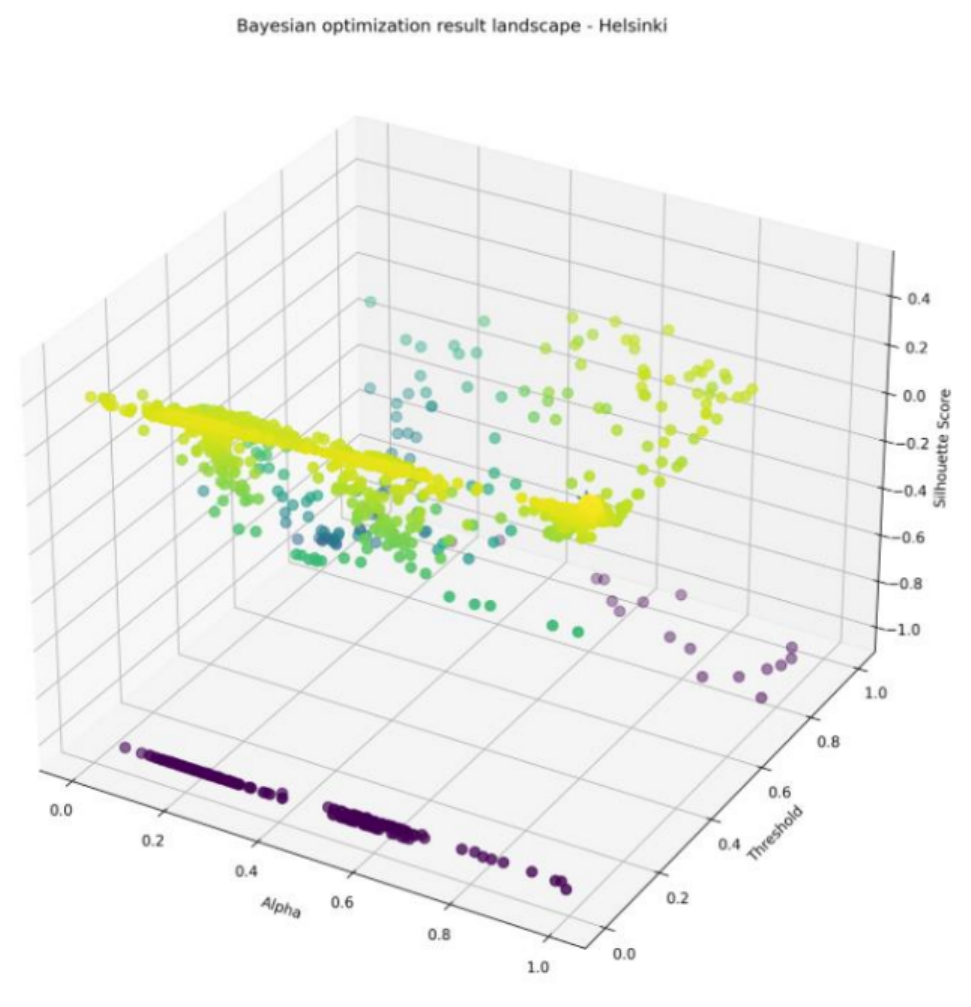}
\caption{Bayesian optimization result for public transport stop clustering in Helsinki.}
\label{figs1}
\end{figure}

As in the picture, the goal was to find the highest point in this 3D space.
For Helsinki, the best clustering parameters were $\alpha=0.9826$, a threshold of 0.0755, and a silhouette score of 0.4748. For example, the final formula is $0.9826 \times distance^{norm}_{ij} + 0.01739 \times (1-namesim_{ij})$, and by applying a threshold of 0.07553, we achieve a silhouette score of 0.4748. In Madrid, the optimal values were $\alpha$ = 0.9991, a threshold of 0.06057, and a silhouette score of 0.06057. The difference in the $\alpha$ values is striking and reflects how the structure of each network influences the clustering process. In Helsinki, especially after we cleaned up the stop name suffixes, hub stops almost always shared the same name, so name-based clustering worked very well. In Madrid, however, stops within the same hub often have different names, which meant that relying only on names was less effective and we needed to put more weight on spatial distance.
In both cities, we found that combining both distance and name similarity gave much better clustering results than using either one alone. This approach consistently led to higher silhouette scores and more meaningful stop clusters.

\subsection{Network creation}

The fundamental approach to constructing public transportation networks was consistent across all cities included in the analysis. Each network was built using publicly available \acrfull{GTFS} data. We modeled the transit system as a directed multigraph, where each node corresponds to a public transport stop and the directed edges represent vehicle connections between them. The weight of each edge is given by the scheduled travel time. In cases where the \acrshort{GTFS} reported a travel time of zero seconds due to an incorrect data record, we substituted a minimum value of 30 seconds to ensure realism. Each edge also stores its associated route id, making it possible to track the specific vehicle line for each connection.
To achieve a fully connected network, we introduced footpaths—walking edges between stops within the same cluster—to represent transfers on foot. Consistent with transit-routing models, these edges have a fixed duration (180 seconds, set in consultation with practitioners) and are labeled ‘walking’ for traceability \cite{delling2015round}.  After constructing the entire network, we extracted the largest strongly connected component to focus our analysis on the fully connected network. For consistency, the network was generated based on the most densely scheduled hour of September 21st, 2020 in the case of Budapest, while in the case of Madrid and Helsinki we used 26th of August, 2022 and 17th of October, 2019 respectively, because these dates were selected based on the data availability.
For reachability analysis, we use a depth-first search (DFS) that explores all feasible paths from a given stop, under a 10-minute (600 seconds) total travel-time budget \cite{marra2020determining}. As the search traverses the network it accumulates edge weights. When transferring between two non-walking routes at the same stop (route ID changes and neither edge is ‘walking’), we add a 180 seconds transfer penalty to simulate vehicle switch time. Transfers via a walking edge (i.e., within-cluster footpaths) incur no extra penalty, as the walking edge already includes the transfer time. If adding the next vehicle edge would exceed the 600 seconds budget, that edge is pruned from the search.
The algorithm records both all feasible paths and the set of stops reachable within the time threshold from each origin. This process is repeated for every stop in the largest connected component, ultimately producing a comprehensive mapping of 10-minute accessibility across the entire network.

\clearpage
\section{Amenity access from OpenStreetMap}
\label{si:sec:amenity_access_from_osm}

The amenities are extracted from \acrfull{OSM}.
In \acrshort{OSM} terminology ''amenity'' is only one of the 5 \acrfull{POI} types: amenity, leisure, office, shop, and tourism.
When ''amenity'' is mentioned in this study, all five types are considered.
The amenity type is one of the five aforementioned \acrshort{POI} types, and the amenity subtype can be something like a restaurant, a school, an atm, a playground, an artwork and so on.
During the data processing a ''category'' is formed from the amenity type and subtype, with a colon as a delimiter, like: amenity:restaurant, amenity:school, amenity:atm, leisure:playground, tourism:artwork.
In the case of Budapest, there are 591 amenity categories in the \acrshort{OSM} data, but we only focus on the 'essential amenities', defined in \cite{abbiasov202415}.

Abbiasov et al. uses \acrfull{NAICS} codes in nine categories: Healthcare, Drugstores, Schools, Religious Organizations, Restaurants, Groceries, Parks, Cultural Institutions, and Services.

In the OpenStreetMap wiki, there are unofficial \acrshort{NAICS} to \acrshort{OSM} tag mapping pages for the 2017 \cite{wiki_naics_2017}, and the 2022 \cite{wiki_naics_2022} versions of the \acrshort{NAICS}.
Note that \cite{abbiasov202415} always uses 6-digit codes, but often the shorter, direct parent can be mapped to \acrshort{OSM} categories.
Table~\ref{tab:essential_categories_by_naics} shows the corresponding 6-digit NAICS codes for the essential amenity categories \cite{abbiasov202415}.
Table~\ref{tab:naics_to_osm} shows how \acrshort{NAICS} codes are mapped to \acrshort{OSM} categories.
Note that \acrshort{NAICS} codes are hierarchical, if a higher category could be mapped to \acrshort{OSM} then the subcategories are omitted, so as the trailing zeros from the codes.

As NAICS code are for business activities, public parks cannot be represented, however, they are integral part of the recreation and should be part of the 15-minute city concept.
So, the categories leisure:park, leisure:playground, and leisure:nature\_reserve are included to the essential amenity category Parks.

There are also some cultural institutions in the \acrshort{OSM} data that are not directly covered by the NAICS mapping, but included to the essential amenity category Cultural institutions, such as: amenity:arts\_centre, tourism:gallery, amenity:theatre, and theatre\_type=concert\_hall.
Optionally amenity:cinema or leisure:stadium could be added as sport facilities mentioned in the description.

\begin{table}[!ht]
    \centering
    \caption{Amenity categories and respective \acrfull{NAICS} codes based on \cite{abbiasov202415}.}
    \begin{tabular}{ll}
    \hline
        \textbf{Category} & \textbf{6-Digit NAICS Code} \\ \hline
        Healthcare & 621111, 621112, 621410, 621492, 621493, 621498, 622110 \\
        Drugstores & 446110 \\
        Schools & 611110 \\
        Religious Organizations & 813110 \\
        Restaurants & 722511, 722513, 722514, 722515 \\
        Groceries & 445110, 445120 \\
        Parks & 712190 \\
        Cultural Institutions & 711190, 711310, 712110 \\
        Services & 491110, 522110, 812111, 812112, 812113, 812320 \\ \hline
    \end{tabular}
    \label{tab:essential_categories_by_naics}
\end{table}

\begin{table}[!ht]
    \centering
    \caption{\acrlong{OSM} categories mapped to \acrfull{NAICS} codes based on \cite{wiki_naics_2017}.}
    \begin{tabularx}{\textwidth}{|l|X|X|}
    \hline
        \textbf{NAICS} & \textbf{} & \textbf{OpenStreepMap category} \\ \hline
        6211 & Offices of Physicians & amenity:doctors or healthcare:doctor \\ \hline
        6214 & Outpatient Care Centers & no equivalent but hospitals should cover \\ \hline
        622 & Hospitals & amenity:hospital or healthcare:hospital \\ \hline
        622110 & General Medical and Surgical Hospitals & ~ \\ \hline
        44611 & Pharmacies and Drug Stores & shop:chemist or amenity:pharmacy \\ \hline
        61111 & Elementary and Secondary Schools & amenity:school or amenity:kindergarten or amenity:preschool \\ \hline
        8131 & Religious Organizations & office:religion or office:parish or amenity:place\_of\_worship \\ \hline
        722511 & Full-Service Restaurants & amenity:restaurant \\ \hline
        722513 & Limited-Service Restaurants & amenity:fast\_food or amenity:food\_court \\ \hline
        722514 & Cafeterias, Grill Buffets, and Buffets & fast\_food:cafeteria \\ \hline
        722515 & Snack and Nonalcoholic Beverage Bars & amenity:cafe or amenity:ice\_cream \\ \hline
        44511 & Supermarkets and Other Grocery (except Convenience) Stores & shop:supermarket or shop:grocery \\ \hline
        44512 & Convenience Stores & shop:convenience \\ \hline
        71219 & Nature Parks and Other Similar Institutions & zoo:wildlife\_park or zoo:safari\_park or tourism:zoo \\ \hline
        711190 & Other Performing Arts Companies & no \acrshort{OSM} equivalent \\ \hline
        711310 & Promoters of Performing Arts, Sports, and Similar Events with Facilities & no \acrshort{OSM} equivalent \\ \hline
        71211 & Museums & tourism:museum \\ \hline
        491 & Postal Service & amenity:post\_office \\ \hline
        522110 & Depository Credit Intermediation & amenity:bank \\ \hline
        812111 & Barber Shops & shop:hairdresser \\ \hline
        812112 & Beauty Salons & shop:beauty or leisure:tanning\_salon \\ \hline
        812113 & Nail Salons & beauty:nails \\ \hline
        8123 & Drycleaning and Laundry Services & shop:laundry or shop:dry\_cleaning \\ \hline
    \end{tabularx}
    \label{tab:naics_to_osm}
\end{table}

\clearpage
\section{Calculating access area by walk}
\label{si:sec:calculating_access_area_by_walk}

Calculating an access area from a point requires a routing engine.
In this study, we used the one called Valhalla 3.2.0 \cite{valhalla320}.
Valhalla processes \acrshort{OSM} data and saves it in its graph format, which can be directly loaded by the pyValhalla package and all routing can still be done efficiently without running the service.

\begin{figure}[th]
\centering
\includegraphics[width=0.75\textwidth]{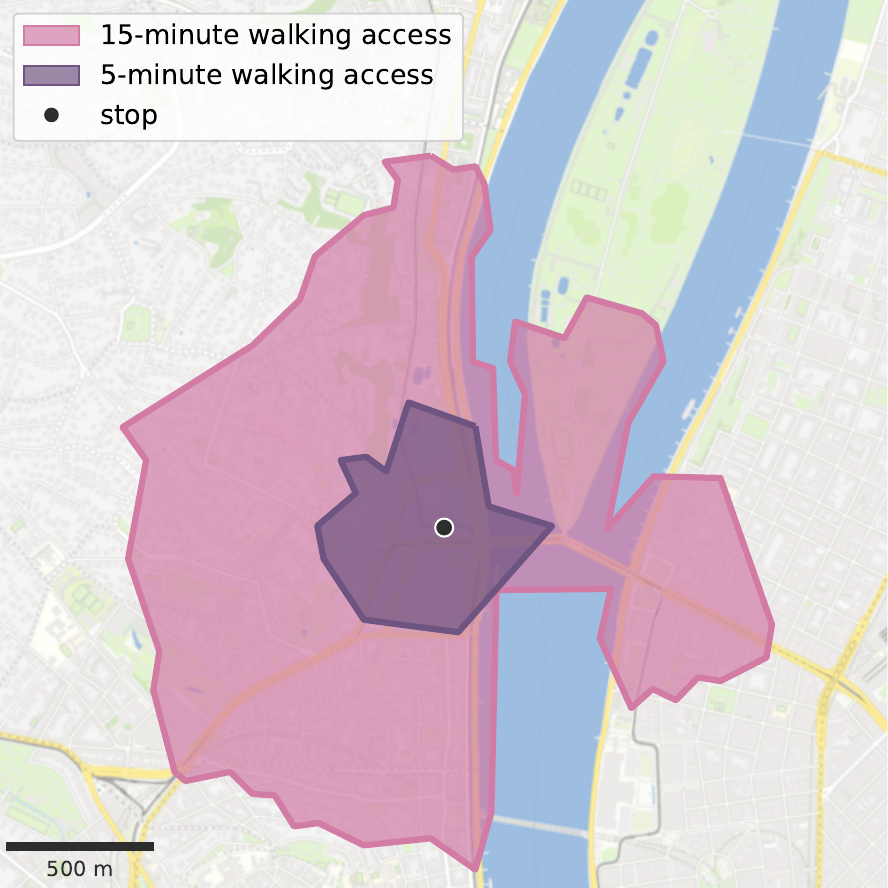}
\caption{The 5- and 15-minute walking areas from the stop called 'Margit híd, budai hídfő'.}
\label{fig:isochrones}
\end{figure}

Valhalla as a routing engine can determine the route between two points in a city with directions, the distance of the route, the trajectory, and time required to take by the given mean of transport like walking, biking, or drivng a car.
It can also determine isochrones, the area accessible from a point in a given time .
The 5- and 15-minute walking area were calculated for every stop of the public transport system (see Figure~\ref{fig:isochrones}).

The 15-minute isochrone determines the area what is possible to reach within 15 minutes by walking. To compare with the accessible area by public transport, we use 5-minute walking areas from the stops, connected with the public transport routing.

\clearpage
\section{Budapest data}
\label{si:sec:budapest_data}

Socio-economic information, including Gini and income groups for Budapest were estimated using two different datasets. The experienced mixing potential is based on the \acrfull{TPDD} we use in the study. This geolocated data comes from Magyar Telekom, a leading Hungarian telecommunications service provider. It provides aggregated population density data at a 100x100 m hexagonal level, based on device traffic at one-hourly intervals, and we use September 21st, 2020 at 3PM. This population density data is derived from the raw geolocated data available to the company and is based on telecommunications events occurring on 2G, 3G and 4G networks on mobile phones and tablets only (e.g. phone calls, text messages, internet usage and movement between cell tower areas). These events are used to locate the area via triangulation. Devices are linked to demographic data derived from anonymized subscriber databases, and therefore we know the number of people in low, middle and high income groups at a given location in the given hour.

\subsection{Real estate prices for Budapest}
On the other hand, we also estimate socio-economic information from residential data using housing prices.
The housing data originates from the \textit{ingatlan.com} real estate website.
The data contains almost \num{62000} property listings within the administrative boundaries of Budapest, including floor space and selling prices from advertisements.
Although prices may not reflect the actual amount paid by the buyer, the order of magnitude should be reasonably accurate, even if there was some bargaining.

The data is from 2018, older than the rest of data used in this study.
 Real estate prices have increased significantly since 2018 in Budapest, but the relative difference between neighborhoods did not change dramatically.
The price of one square meter was calculated from the floor space and the advertising price. In this way, the price level of two different estates in two very different parts of the city can be compared.
This data was used before in \cite{pinter2021evaluating} to proxy socio-economic status.

\begin{figure}[ht]
    \centering
    \begin{subfigure}{0.57\linewidth}
        \includegraphics[width=\linewidth]{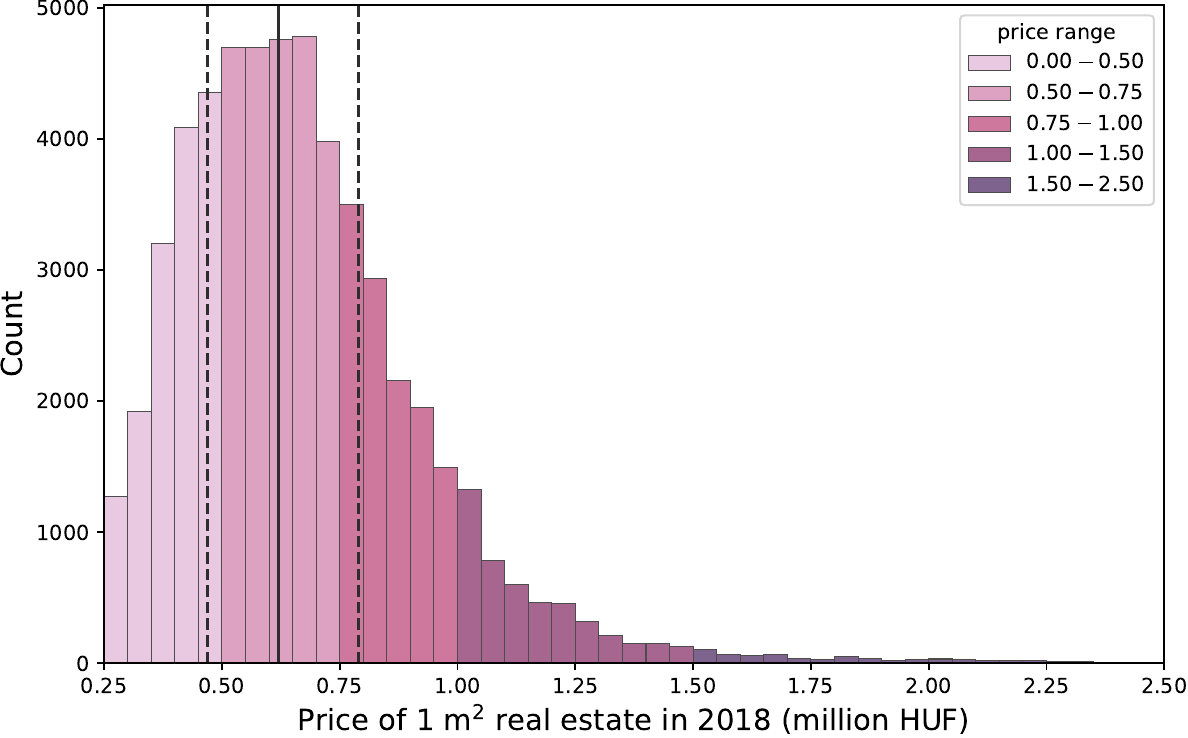}
        \caption{}
        \label{fig:property_price_distribution}
    \end{subfigure}
    \begin{subfigure}{0.42\linewidth}
        \includegraphics[width=\linewidth]{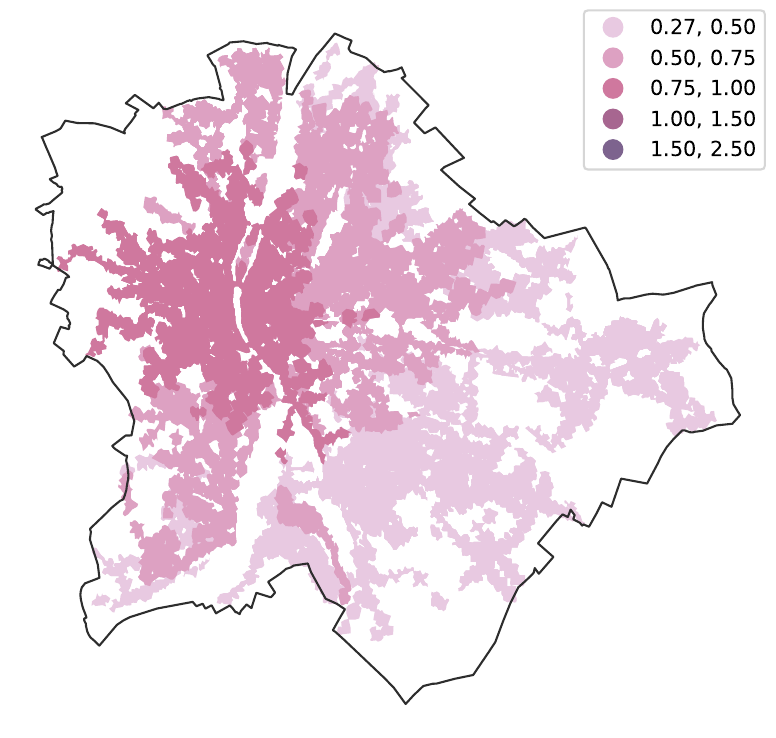}
        \caption{}
        \label{fig:property_price_spatial_distribution}
    \end{subfigure}

    \caption{Property price distribution (\textbf{\subref{fig:property_price_distribution}}), and the average property price in a 5-minute access area of a stop (\textbf{\subref{fig:property_price_spatial_distribution}}).}
    \label{fig:property_price}
\end{figure}

\num {57549} listings are within the 5-minute walking range of the stops.
For each 5- and 15-minute access area of a stop, the mean property price is calculated (Figure~\ref{fig:property_price_spatial_distribution}).

\clearpage
\section{Socio-economic measure}
\label{si:sec:socioeconomic_measure}

We use Olivia Guest's Python Gini implementation \cite{guest2016gini_code} to estimate Gini in Helsinki and Budapest, which implements the formula (\ref{eq:gini}).

\begin{equation}
    \label{eq:gini}
    G = \frac{\sum^n_{i=1}(2i-n-1)x_i}{n\sum^n_{i=n}x_i}
\end{equation}

\subsection{Helsinki}

To estimate the Gini coefficient, we first derive representative income levels for three socio-economic groups using national gross income data organized by decile. We define low-income households as those in deciles I–II and high-income households as those in deciles IX–X. We then take the mean gross income within each group. The fifth decile serves as the national median income reference. Next, we compute the relative deviation of the low- and high-income group means from the national median. We apply these relative deviations to each postal code's local median household income to produce postal code–specific representative incomes for low- and high-income households, respectively. The local median itself serves as the representative income for middle-income households. Finally, for each postal code, we construct a weighted income array by repeating each representative income value according to the number of households in the corresponding income category. We then apply the Gini formula (\ref{eq:gini}) to this distribution.

We now have Gini coefficients estimated and and median incomes available from Paavo database by Statistics Finland \cite{statfi2026paavo} at the postal code level, we translate these estimates to the walking and multimodal polygons using an area-weighted aggregation. For each polygon, we use a spatial join to identify all intersecting postal codes and compute the intersection area between the catchment and each postal code. We then weight each postal code's contribution by the share of its area that falls within the catchment, ensuring that postal codes with greater overlap have proportionally more influence. We then compute the weighted Gini coefficient and weighted median income for each catchment polygon as the weighted mean across all intersecting postal codes.
To differentiate between neighborhoods by residential status in the regressions, we divide our data at the median income in each walking polygon, to low and high status areas.

\subsection{Budapest}
In Budapest, average income data are unavailable at fine spatial scales. To measure the experienced social mixing, we use the gross mean monthly income in 2024 for the smallest available geographic unit: the district. Since districts are less granular than postal codes, we calculate the mean of gross income across all districts intersecting each catchment polygon. This finer spatial discretisation reduces the boundary effect of large administrative units (such as postal codes in Helsinki or census tracts in Madrid) and provides a more spatially precise income assignment across the catchment area. To construct the income distribution, we use the same relative deviation approach as in Helsinki, employing national income decile data from the Hungarian Central Statistical Office (KSH). We define low income as deciles I–II and high income as deciles IX–X. Note that the decile data are from 2020, the most recent year available, while the district-level income figures are from 2024. We apply the relative deviations to the catchment-level income to produce representative low- and high-income estimates. We then construct a weighted income array using the number of residents by income group at their home location, derived from Telekom mobile network data.

To differentiate between neighborhoods by experienced income, we divide our data based on the ratio of low income people at the 15-minute walking area, and areas with higher ratio of low income population than the median will be labeled low, those with lower ratio will be labeled high income.

We use a different approach for the residential Gini. Instead of estimating income distributions, we use property listing data from ingatlan.com, which provides square meter prices at the house level. Because this data captures individual properties directly, we can apply the Gini formula (\ref{eq:gini}) to the distribution of square meter prices within each catchment polygon without further estimation.

To differentiate between neighborhoods by residential status, we divide our data at the mean house price in each walking polygon, to low and high status areas.

\subsection{Madrid}
In Madrid, INE provides Gini coefficients and median income directly at the census tract level because it has access to granular, household-level data. Therefore, we use these coefficients directly rather than estimating them. This contrasts with the estimation approach used for Helsinki and Budapest, where income distributions are reconstructed from grouped data. Since grouping households into broad income categories compresses the true income distribution, this likely results in a systematic underestimation of inequality in those two cities compared to Madrid. However, we apply the same area-weighted aggregation as in Helsinki to translate census tract level values to walking and multimodal polygons.

\clearpage
\section{Ellipticity}
\label{si:sec:ellipticity}

\begin{figure}[ht]
    \includegraphics[width=\linewidth]{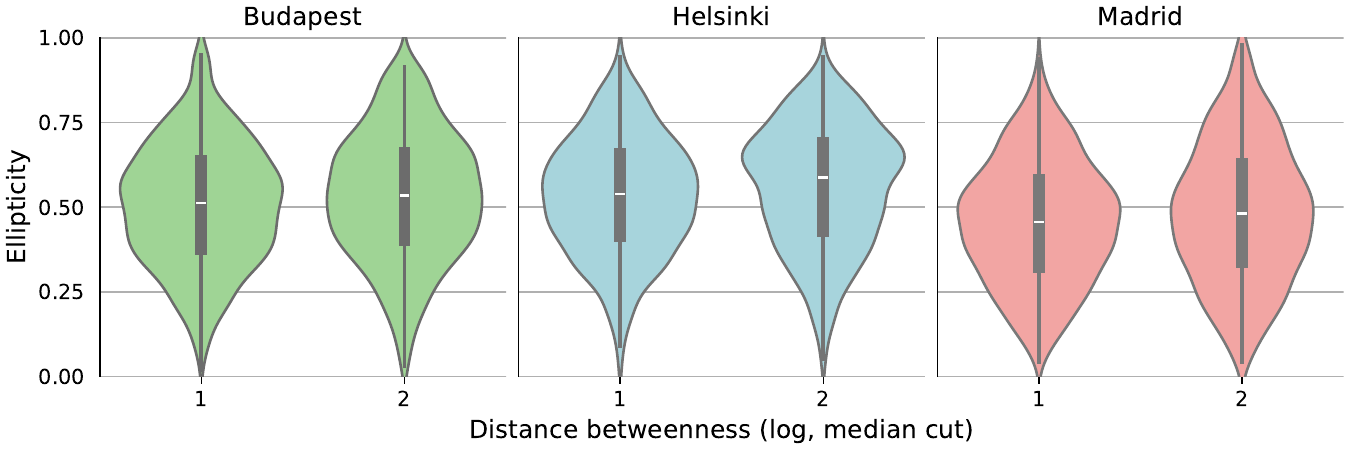}
    \caption{The mean of $E_p$ is higher in peripheral neighborhoods than in the center (defined by median distance from center of gravity) but have similarly wide distributions in all three cities.}
    \label{fig:si:ellipticity_by_distance_betweenness}
\end{figure}

Figure ~\ref{fig:si:ellipticity_by_distance_betweenness} shows that the shape of public transport is different closer and further from the center, areas further from the center have on average more elliptic shapes. However, the distribution of $E_p$ signals a large variety of even and uneven multimodal access in urban peripheries as well (Figure \ref*{fig:fig2}B of the main text).

\clearpage
\section{Descriptive statistics and correlations}
\label{si:sec:descriptive_stats}

\begin{figure}[htbp]
\centering
\includegraphics[width=\linewidth]{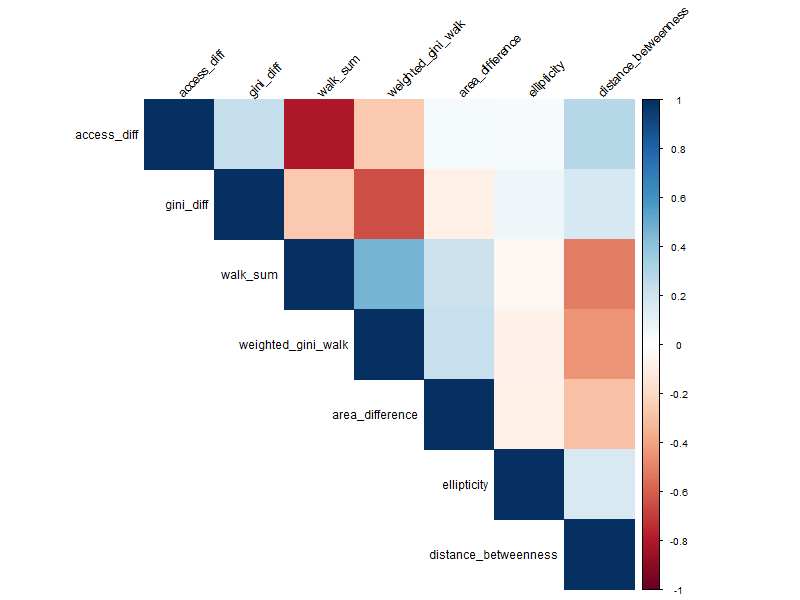}
\caption{Correlation structure of variables in Helsinki.}
\label{fig:corr_helsinki}
\end{figure}

\begin{figure}[htbp]
\centering
\includegraphics[width=\linewidth]{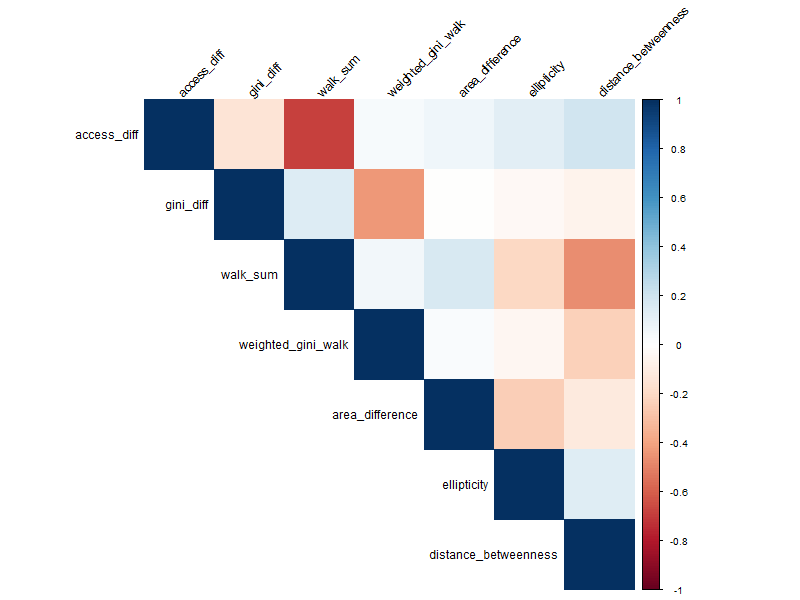}
\caption{Correlation structure of variables in Madrid.}
\label{fig:corr_madrid}
\end{figure}

\begin{figure}[htbp]
\centering
\includegraphics[width=\linewidth]{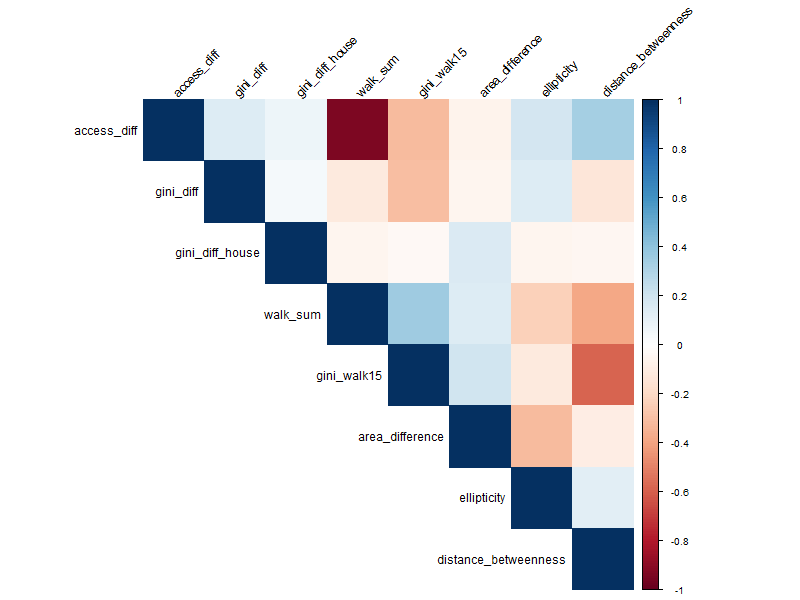}
\caption{Correlation structure of variables in Budapest.}
\label{fig:corr_bp}
\end{figure}

In this Supplementary section, we document the descriptive statistics of main variables in Helsinki, Madrid, and Budapest in Tables S3-S5. Then, we illustrate the Pearson correlation coefficients between these variables in Figures~\ref{fig:corr_helsinki}-~\ref{fig:corr_madrid}-~\ref{fig:corr_bp}.

\begin{table}[!htbp] \centering 
  \caption{Summary Statistics for Helsinki} 
\begin{tabular}{@{\extracolsep{5pt}}lccccccc} 
\\[-1.8ex]\hline 
\hline \\[-1.8ex] 
Statistic & \multicolumn{1}{c}{Min} & \multicolumn{1}{c}{Pctl(25)} & \multicolumn{1}{c}{Median} & \multicolumn{1}{c}{Mean} & \multicolumn{1}{c}{Pctl(75)} & \multicolumn{1}{c}{Max} & \multicolumn{1}{c}{St. Dev.} \\ 
\hline \\[-1.8ex] 
access\_diff & $-$4 & 0 & 1 & 1.212 & 2 & 8 & 1.349 \\ 
gini\_diff & $-$0.056 & $-$0.003 & 0.0004 & 0.002 & 0.005 & 0.074 & 0.010 \\ 
walk\_sum & 1 & 6 & 8 & 7.160 & 8 & 9 & 1.652 \\ 
weighted\_gini\_walk & 0.231 & 0.301 & 0.311 & 0.307 & 0.319 & 0.337 & 0.019 \\ 
area\_difference & $-$1.687 & 1.579 & 3.434 & 4.348 & 6.086 & 21.887 & 3.771 \\ 
ellipticity & 0.026 & 0.423 & 0.562 & 0.550 & 0.678 & 0.945 & 0.175 \\ 
distance\_betweenness & 0.229 & 4.019 & 6.141 & 6.485 & 8.721 & 19.908 & 3.248 \\ 
\hline \\[-1.8ex] 
\end{tabular} 
\end{table}

\begin{table}[!htbp] \centering 
  \caption{Summary Statistics for Madrid} 
\begin{tabular}{@{\extracolsep{5pt}}lccccccc} 
\\[-1.8ex]\hline 
\hline \\[-1.8ex] 
Statistic & \multicolumn{1}{c}{Min} & \multicolumn{1}{c}{Pctl(25)} & \multicolumn{1}{c}{Median} & \multicolumn{1}{c}{Mean} & \multicolumn{1}{c}{Pctl(75)} & \multicolumn{1}{c}{Max} & \multicolumn{1}{c}{St. Dev.} \\ 
\hline \\[-1.8ex] 
access\_diff & $-$4 & 0 & 0 & 0.219 & 0 & 8 & 0.811 \\ 
gini\_diff & $-$11.428 & $-$0.500 & 0.052 & $-$0.061 & 0.565 & 7.814 & 1.306 \\ 
walk\_sum & 1 & 8 & 9 & 8.390 & 9 & 9 & 1.101 \\ 
weighted\_gini\_walk & 20.494 & 29.784 & 31.658 & 32.444 & 34.353 & 44.400 & 3.494 \\ 
area\_difference & $-$2.570 & 0.073 & 0.982 & 1.192 & 2.074 & 12.137 & 1.593 \\ 
ellipticity & 0.008 & 0.337 & 0.480 & 0.481 & 0.623 & 0.986 & 0.195 \\ 
distance\_betweenness & 0.020 & 3.861 & 5.821 & 5.970 & 7.918 & 16.529 & 2.808 \\ 
\hline \\[-1.8ex] 
\end{tabular} 
\end{table}

\begin{table}[!htbp] \centering 
  \caption{Summary Statistics for Budapest} 
\begin{tabular}{@{\extracolsep{5pt}}lccccccc} 
\\[-1.8ex]\hline 
\hline \\[-1.8ex] 
Statistic & \multicolumn{1}{c}{Min} & \multicolumn{1}{c}{Pctl(25)} & \multicolumn{1}{c}{Median} & \multicolumn{1}{c}{Mean} & \multicolumn{1}{c}{Pctl(75)} & \multicolumn{1}{c}{Max} & \multicolumn{1}{c}{St. Dev.} \\ 
\hline \\[-1.8ex] 
access\_diff & $-$2 & 0 & 0 & 0.780 & 1 & 8 & 1.441 \\ 
gini\_diff & $-$0.094 & $-$0.003 & 0.004 & 0.008 & 0.015 & 0.169 & 0.020 \\ 
gini\_diff\_house & $-$0.215 & $-$0.016 & 0.002 & $-$0.001 & 0.021 & 0.181 & 0.045 \\ 
walk\_sum & 1 & 8 & 9 & 8.070 & 9 & 9 & 1.554 \\ 
gini\_walk15 & 0.000 & 0.119 & 0.130 & 0.142 & 0.150 & 0.272 & 0.039 \\ 
area\_difference & $-$1.596 & 1.659 & 3.160 & 3.590 & 5.092 & 18.770 & 2.742 \\ 
ellipticity & 0.020 & 0.390 & 0.523 & 0.519 & 0.655 & 0.950 & 0.184 \\ 
distance\_betweenness & 0.185 & 5.011 & 7.438 & 7.653 & 10.016 & 19.528 & 3.628 \\ 
\hline \\[-1.8ex] 
\end{tabular} 
\end{table}

\clearpage

\subsection{Regression tables}

This Supplementary section documents the regression results that are illustrated in Figures 3 and 4 in the main text. Tables S6-S7-S8 contain coefficients and standard errors from the estimation that does not include interaction terms. The coefficients of Area Difference and Ellipticity in Figures 3 and 4 refer to these values. Tables S9-S10-S11 contain coefficients and standard errors from the estimation that does include the interaction term between Ellipticity and Distance from the Center. The marginal plots in Figures 3 and 4 are calculated from these regressions.

{\small

\begin{sidewaystable}
\centering 
  \caption{Results of the linear regression without interaction terms and socio-economic decomposition. Coefficients of Area Difference and Ellipticity are depicted in Figures 3 and 4 of the main text.} 
\begin{tabular}{@{\extracolsep{2pt}}lccccccc} 
\\[-1.8ex]\hline 
\hline \\[-1.8ex] 
 & Helsinki & Madrid & BP & Helsinki & Madrid & BP residential & BP experienced \\ 
  & Access & Access & Access & Gini & Gini & Gini & Gini \\ 
\\[-1.8ex] & (1) & (2) & (3) & (4) & (5) & (6) & (7)\\ 
\hline \\[-1.8ex] 
 walk\_sum & $-$0.920$^{***}$ & $-$0.780$^{***}$ & $-$0.965$^{***}$ &  &  &  &  \\ 
  & (0.018) & (0.011) & (0.009) &  &  &  &  \\ 
  & & & & & & & \\ 
 weighted\_gini\_walk &  &  &  & $-$0.706$^{***}$ & $-$0.474$^{***}$ &  &  \\ 
  &  &  &  & (0.026) & (0.012) &  &  \\ 
  & & & & & & & \\ 
 gini\_walk15 &  &  &  &  &  & $-$0.006$^{***}$ & $-$0.590$^{***}$ \\ 
  &  &  &  &  &  & (0.001) & (0.026) \\ 
  & & & & & & & \\ 
 area\_difference & 0.194$^{***}$ & 0.178$^{***}$ & 0.076$^{***}$ & 0.032 & $-$0.017 & 0.008$^{***}$ & 0.062$^{***}$ \\ 
  & (0.017) & (0.010) & (0.008) & (0.024) & (0.013) & (0.001) & (0.022) \\ 
  & & & & & & & \\ 
 ellipticity & 0.036$^{**}$ & 0.022$^{**}$ & $-$0.016$^{*}$ & 0.030 & $-$0.033$^{***}$ & $-$0.000 & 0.156$^{***}$ \\ 
  & (0.016) & (0.010) & (0.009) & (0.023) & (0.013) & (0.001) & (0.022) \\ 
  & & & & & & & \\ 
 distance\_betweenness & $-$0.144$^{***}$ & $-$0.143$^{***}$ & $-$0.029$^{***}$ & $-$0.132$^{***}$ & $-$0.172$^{***}$ & $-$0.005$^{***}$ & $-$0.499$^{***}$ \\ 
  & (0.019) & (0.011) & (0.009) & (0.027) & (0.013) & (0.001) & (0.026) \\ 
  & & & & & & & \\ 
 Constant & 0.007 & 0.011 & $-$0.000 & 0.004 & $-$0.003 & $-$0.001 & $-$0.000 \\ 
  & (0.016) & (0.009) & (0.008) & (0.023) & (0.012) & (0.001) & (0.021) \\ 
  & & & & & & & \\ 
Observations & 1,137 & 5,211 & 1,726 & 1,087 & 5,211 & 1,726 & 1,726 \\ 
R$^{2}$ & 0.715 & 0.515 & 0.889 & 0.427 & 0.221 & 0.037 & 0.267 \\ 
\hline \\[-1.8ex] 
\textit{Notes:} & \multicolumn{7}{l}{$^{***}$Significant at the 1 percent level.} \\ 
 & \multicolumn{7}{l}{$^{**}$Significant at the 5 percent level.} \\ 
 & \multicolumn{7}{l}{$^{*}$Significant at the 10 percent level.} \\ 
\end{tabular} 
\end{sidewaystable}

\begin{sidewaystable}
\centering 
  \caption{Results of the linear regression without interaction terms for the low socio-economic status neighborhoods. Coefficients of Area Difference and Ellipticity are depicted in Figures 3 and 4 of the main text.} 
\begin{tabular}{@{\extracolsep{5pt}}lccccccc} 
\\[-1.8ex]\hline 
\hline \\[-1.8ex] 
 & Helsinki & Madrid & BP & Helsinki & Madrid & BP residential & BP experienced \\ 
  & Access & Access & Access & Gini & Gini & Gini & Gini \\ 
\\[-1.8ex] & (1) & (2) & (3) & (4) & (5) & (6) & (7)\\ 
\hline \\[-1.8ex] 
 walk\_sum & $-$1.019$^{***}$ & $-$0.704$^{***}$ & $-$0.989$^{***}$ &  &  &  &  \\ 
  & (0.024) & (0.015) & (0.011) &  &  &  &  \\ 
  & & & & & & & \\ 
 weighted\_gini\_walk &  &  &  & $-$0.740$^{***}$ & $-$0.514$^{***}$ &  &  \\ 
  &  &  &  & (0.043) & (0.023) &  &  \\ 
  & & & & & & & \\ 
 gini\_walk15 &  &  &  &  &  & $-$0.192$^{***}$ & $-$0.498$^{***}$ \\ 
  &  &  &  &  &  & (0.069) & (0.039) \\ 
  & & & & & & & \\ 
 area\_difference & 0.167$^{***}$ & 0.184$^{***}$ & 0.103$^{***}$ & $-$0.037$^{**}$ & 0.046$^{***}$ & 0.074$^{*}$ & 0.044 \\ 
  & (0.018) & (0.013) & (0.013) & (0.018) & (0.011) & (0.039) & (0.029) \\ 
  & & & & & & & \\ 
 ellipticity & 0.086$^{***}$ & 0.033$^{**}$ & $-$0.036$^{***}$ & $-$0.012 & $-$0.022$^{*}$ & 0.091$^{**}$ & 0.208$^{***}$ \\ 
  & (0.019) & (0.014) & (0.013) & (0.019) & (0.012) & (0.036) & (0.033) \\ 
  & & & & & & & \\ 
 distance\_betweenness & $-$0.090$^{***}$ & $-$0.050$^{***}$ & $-$0.044$^{***}$ & $-$0.119$^{***}$ & $-$0.108$^{***}$ & $-$0.153$^{***}$ & $-$0.596$^{***}$ \\ 
  & (0.021) & (0.018) & (0.014) & (0.028) & (0.017) & (0.040) & (0.062) \\ 
  & & & & & & & \\ 
 Constant & 0.078$^{***}$ & 0.011 & 0.023 & 0.061$^{**}$ & $-$0.164$^{***}$ & 0.044 & $-$0.059 \\ 
  & (0.018) & (0.014) & (0.014) & (0.026) & (0.016) & (0.049) & (0.040) \\ 
  & & & & & & & \\ 
Observations & 569 & 2,603 & 862 & 525 & 2,603 & 862 & 863 \\ 
R$^{2}$ & 0.777 & 0.474 & 0.907 & 0.446 & 0.190 & 0.029 & 0.206 \\ 
\hline \\[-1.8ex] 
\textit{Notes:} & \multicolumn{7}{l}{$^{***}$Significant at the 1 percent level.} \\ 
 & \multicolumn{7}{l}{$^{**}$Significant at the 5 percent level.} \\ 
 & \multicolumn{7}{l}{$^{*}$Significant at the 10 percent level.} \\ 
\end{tabular} 
\end{sidewaystable}

\begin{sidewaystable}
\centering 
  \caption{Results of the linear regression without interaction terms for the high socio-economic status neighborhoods. Coefficients of Area Difference and Ellipticity are depicted in Figures 3 and 4 of the main text.} 
\begin{tabular}{@{\extracolsep{5pt}}lccccccc} 
\\[-1.8ex]\hline 
\hline \\[-1.8ex] 
 & Helsinki & Madrid & BP & Helsinki & Madrid & BP residential & BP experienced \\ 
  & Access & Access & Access & Gini & Gini & Gini & Gini \\ 
\\[-1.8ex] & (1) & (2) & (3) & (4) & (5) & (6) & (7)\\  
\hline \\[-1.8ex] 
 walk\_sum & $-$0.920$^{***}$ & $-$0.869$^{***}$ & $-$0.927$^{***}$ &  &  &  &  \\ 
  & (0.028) & (0.016) & (0.015) &  &  &  &  \\ 
  & & & & & & & \\ 
 weighted\_gini\_walk &  &  &  & $-$0.741$^{***}$ & $-$0.593$^{***}$ &  &  \\ 
  &  &  &  & (0.041) & (0.019) &  &  \\ 
  & & & & & & & \\ 
 gini\_walk15 &  &  &  &  &  & $-$0.141$^{***}$ & $-$1.312$^{***}$ \\ 
  &  &  &  &  &  & (0.042) & (0.049) \\ 
  & & & & & & & \\ 
 area\_difference & 0.222$^{***}$ & 0.153$^{***}$ & 0.043$^{***}$ & 0.123$^{***}$ & $-$0.131$^{***}$ & 0.229$^{***}$ & 0.049 \\ 
  & (0.028) & (0.015) & (0.011) & (0.048) & (0.023) & (0.034) & (0.030) \\ 
  & & & & & & & \\ 
 ellipticity & $-$0.017 & 0.022 & $-$0.006 & 0.075$^{*}$ & 0.028 & $-$0.079$^{**}$ & 0.127$^{***}$ \\ 
  & (0.025) & (0.014) & (0.012) & (0.041) & (0.021) & (0.035) & (0.027) \\ 
  & & & & & & & \\ 
 distance\_betweenness & $-$0.236$^{***}$ & $-$0.223$^{***}$ & $-$0.054$^{***}$ & $-$0.164$^{***}$ & $-$0.308$^{***}$ & $-$0.119$^{*}$ & $-$0.313$^{***}$ \\ 
  & (0.033) & (0.014) & (0.018) & (0.046) & (0.019) & (0.070) & (0.030) \\ 
  & & & & & & & \\ 
 Constant & $-$0.023 & 0.033$^{**}$ & $-$0.045$^{***}$ & $-$0.028 & 0.237$^{***}$ & $-$0.038 & $-$0.459$^{***}$ \\ 
  & (0.025) & (0.014) & (0.015) & (0.044) & (0.023) & (0.047) & (0.039) \\ 
  & & & & & & & \\ 
Observations & 567 & 2,607 & 861 & 561 & 2,607 & 861 & 863 \\ 
R$^{2}$ & 0.702 & 0.563 & 0.845 & 0.388 & 0.282 & 0.080 & 0.495 \\ 
\hline \\[-1.8ex] 
\textit{Notes:} & \multicolumn{7}{l}{$^{***}$Significant at the 1 percent level.} \\ 
 & \multicolumn{7}{l}{$^{**}$Significant at the 5 percent level.} \\ 
 & \multicolumn{7}{l}{$^{*}$Significant at the 10 percent level.} \\ 
\end{tabular} 
\end{sidewaystable}

\begin{sidewaystable}
\centering 
  \caption{Results of the linear regression with interaction terms and without socio-economic decomposition. Coefficients of Area Difference and Ellipticity are depicted in Figures 3 and 4 of the main text.} 
\begin{tabular}{@{\extracolsep{5pt}}lccccccc} 
\\[-1.8ex]\hline 
\hline \\[-1.8ex] 
 & Helsinki & Madrid & BP & Helsinki & Madrid & BP residential & BP experienced \\ 
  & Access & Access & Access & Gini & Gini & Gini & Gini \\ 
\\[-1.8ex] & (1) & (2) & (3) & (4) & (5) & (6) & (7)\\ 
\hline \\[-1.8ex] 
 walk\_sum & $-$0.601$^{***}$ & $-$0.518$^{***}$ & $-$0.843$^{***}$ &  &  &  &  \\ 
  & (0.025) & (0.009) & (0.013) &  &  &  &  \\ 
  & & & & & & & \\ 
 weighted\_gini\_walk &  &  &  & $-$0.988$^{***}$ & 1.864$^{***}$ &  &  \\ 
  &  &  &  & (0.000) & (0.017) &  &  \\ 
  & & & & & & & \\ 
 gini\_walk15 &  &  &  &  &  & $-$0.007$^{***}$ & $-$0.973$^{***}$ \\ 
  &  &  &  &  &  & (0.001) & (0.001) \\ 
  & & & & & & & \\ 
 area\_difference & 0.259$^{***}$ & 0.155$^{***}$ & 0.123$^{***}$ & 0.000 & $-$0.031$^{*}$ & 0.008$^{***}$ & 0.001$^{***}$ \\ 
  & (0.023) & (0.008) & (0.012) & (0.000) & (0.017) & (0.001) & (0.000) \\ 
  & & & & & & & \\ 
 ellipticity & 0.055$^{**}$ & 0.026$^{***}$ & $-$0.010 & 0.000$^{*}$ & $-$0.055$^{***}$ & 0.000 & 0.003$^{***}$ \\ 
  & (0.022) & (0.008) & (0.012) & (0.000) & (0.017) & (0.001) & (0.000) \\ 
  & & & & & & & \\ 
 distance\_betweenness & $-$0.195$^{***}$ & $-$0.119$^{***}$ & $-$0.051$^{***}$ & $-$0.002$^{***}$ & $-$0.236$^{***}$ & $-$0.006$^{***}$ & $-$0.010$^{***}$ \\ 
  & (0.026) & (0.009) & (0.012) & (0.000) & (0.017) & (0.001) & (0.001) \\ 
  & & & & & & & \\ 
 area\_difference:ellipticity & 0.030 & 0.059$^{***}$ & 0.079$^{***}$ & 0.001$^{***}$ & $-$0.033$^{**}$ & 0.003$^{***}$ & $-$0.000 \\ 
  & (0.025) & (0.008) & (0.012) & (0.000) & (0.016) & (0.001) & (0.000) \\ 
  & & & & & & & \\ 
 ellipticity:distance\_betweenness & $-$0.012 & $-$0.011 & $-$0.039$^{***}$ & 0.001$^{***}$ & 0.057$^{***}$ & 0.002$^{*}$ & $-$0.001 \\ 
  & (0.024) & (0.007) & (0.013) & (0.000) & (0.015) & (0.001) & (0.000) \\ 
  & & & & & & & \\ 
 Constant & 8.397$^{***}$ & 8.618$^{***}$ & 8.880$^{***}$ & 0.310$^{***}$ & 32.366$^{***}$ & $-$0.000 & 0.150$^{***}$ \\ 
  & (0.022) & (0.008) & (0.012) & (0.000) & (0.016) & (0.001) & (0.000) \\ 
  & & & & & & & \\ 
Observations & 1,137 & 5,211 & 1,726 & 1,087 & 5,211 & 1,726 & 1,726 \\ 
R$^{2}$ & 0.378 & 0.412 & 0.747 & 1.000 & 0.740 & 0.045 & 1.000 \\ 
\hline \\[-1.8ex] 
\textit{Notes:} & \multicolumn{7}{l}{$^{***}$Significant at the 1 percent level.} \\ 
 & \multicolumn{7}{l}{$^{**}$Significant at the 5 percent level.} \\ 
 & \multicolumn{7}{l}{$^{*}$Significant at the 10 percent level.} \\ 
\end{tabular} 
\end{sidewaystable}

\begin{sidewaystable}
\centering 
  \caption{Results of the linear regression with interaction terms for the low socio-economic status neighborhoods. Coefficients of Area Difference and Ellipticity are depicted in Figures 3 and 4 of the main text.} 
\begin{tabular}{@{\extracolsep{5pt}}lccccccc} 
\\[-1.8ex]\hline 
\hline \\[-1.8ex] 
 & Helsinki & Madrid & BP & Helsinki & Madrid & BP residential & BP experienced \\ 
  & Access & Access & Access & Gini & Gini & Gini & Gini \\ 
\\[-1.8ex] & (1) & (2) & (3) & (4) & (5) & (6) & (7)\\ 
\hline \\[-1.8ex] 
 walk\_sum & $-$0.740$^{***}$ & $-$0.454$^{***}$ & $-$0.870$^{***}$ &  &  &  &  \\ 
  & (0.032) & (0.013) & (0.016) &  &  &  &  \\ 
  & & & & & & & \\ 
 weighted\_gini\_walk &  &  &  & $-$0.988$^{***}$ & 1.823$^{***}$ &  &  \\ 
  &  &  &  & (0.000) & (0.031) &  &  \\ 
  & & & & & & & \\ 
 gini\_walk15 &  &  &  &  &  & $-$0.009$^{***}$ & $-$0.971$^{***}$ \\ 
  &  &  &  &  &  & (0.003) & (0.001) \\ 
  & & & & & & & \\ 
 area\_difference & 0.230$^{***}$ & 0.162$^{***}$ & 0.149$^{***}$ & $-$0.000$^{*}$ & 0.056$^{***}$ & 0.003$^{*}$ & 0.001 \\ 
  & (0.025) & (0.011) & (0.019) & (0.000) & (0.015) & (0.002) & (0.001) \\ 
  & & & & & & & \\ 
 ellipticity & 0.104$^{***}$ & 0.032$^{***}$ & $-$0.037$^{*}$ & $-$0.000 & $-$0.033$^{**}$ & 0.002 & 0.003$^{***}$ \\ 
  & (0.026) & (0.012) & (0.022) & (0.000) & (0.016) & (0.002) & (0.001) \\ 
  & & & & & & & \\ 
 distance\_betweenness & $-$0.115$^{***}$ & $-$0.027$^{*}$ & $-$0.069$^{***}$ & $-$0.001$^{***}$ & $-$0.139$^{***}$ & $-$0.008$^{***}$ & $-$0.013$^{***}$ \\ 
  & (0.029) & (0.015) & (0.020) & (0.000) & (0.024) & (0.002) & (0.001) \\ 
  & & & & & & & \\ 
 area\_difference:ellipticity & $-$0.080$^{***}$ & 0.061$^{***}$ & 0.142$^{***}$ & $-$0.000 & $-$0.019 & 0.006$^{***}$ & $-$0.001 \\ 
  & (0.030) & (0.010) & (0.019) & (0.000) & (0.013) & (0.002) & (0.001) \\ 
  & & & & & & & \\ 
 ellipticity:distance\_betweenness & $-$0.052$^{*}$ & $-$0.019 & $-$0.033 & 0.000 & $-$0.026 & 0.003 & $-$0.002$^{**}$ \\ 
  & (0.030) & (0.013) & (0.023) & (0.000) & (0.017) & (0.002) & (0.001) \\ 
  & & & & & & & \\ 
 Constant & 8.499$^{***}$ & 8.623$^{***}$ & 8.920$^{***}$ & 0.310$^{***}$ & 32.162$^{***}$ & 0.002 & 0.148$^{***}$ \\ 
  & (0.026) & (0.012) & (0.021) & (0.000) & (0.022) & (0.002) & (0.001) \\ 
  & & & & & & & \\ 
Observations & 569 & 2,603 & 862 & 525 & 2,603 & 862 & 863 \\ 
R$^{2}$ & 0.523 & 0.370 & 0.797 & 1.000 & 0.696 & 0.046 & 1.000 \\ 
\hline \\[-1.8ex] 
\textit{Notes:} & \multicolumn{7}{l}{$^{***}$Significant at the 1 percent level.} \\ 
 & \multicolumn{7}{l}{$^{**}$Significant at the 5 percent level.} \\ 
 & \multicolumn{7}{l}{$^{*}$Significant at the 10 percent level.} \\ 
\end{tabular} 
\end{sidewaystable}

\begin{sidewaystable}
\centering 
  \caption{Results of the linear regression with interaction terms for the high socio-economic status neighborhoods. Coefficients of Area Difference and Ellipticity are depicted in Figures 3 and 4 of the main text.} 
\begin{tabular}{@{\extracolsep{5pt}}lccccccc} 
\\[-1.8ex]\hline 
\hline \\[-1.8ex] 
 & Helsinki & Madrid & BP & Helsinki & Madrid & BP residential & BP experienced \\ 
  & Access & Access & Access & Gini & Gini & Gini & Gini \\ 
\\[-1.8ex] & (1) & (2) & (3) & (4) & (5) & (6) & (7)\\ 
\hline \\[-1.8ex] 
 walk\_sum & $-$0.602$^{***}$ & $-$0.599$^{***}$ & $-$0.793$^{***}$ &  &  &  &  \\ 
  & (0.038) & (0.013) & (0.022) &  &  &  &  \\ 
  & & & & & & & \\ 
 weighted\_gini\_walk &  &  &  & $-$0.988$^{***}$ & 1.697$^{***}$ &  &  \\ 
  &  &  &  & (0.000) & (0.026) &  &  \\ 
  & & & & & & & \\ 
 gini\_walk15 &  &  &  &  &  & $-$0.007$^{***}$ & $-$0.988$^{***}$ \\ 
  &  &  &  &  &  & (0.002) & (0.001) \\ 
  & & & & & & & \\ 
 area\_difference & 0.291$^{***}$ & 0.132$^{***}$ & 0.069$^{***}$ & 0.001$^{*}$ & $-$0.183$^{***}$ & 0.010$^{***}$ & 0.001 \\ 
  & (0.038) & (0.012) & (0.017) & (0.000) & (0.031) & (0.002) & (0.001) \\ 
  & & & & & & & \\ 
 ellipticity & 0.005 & 0.032$^{***}$ & $-$0.021 & 0.001$^{**}$ & 0.007 & $-$0.006$^{***}$ & 0.004$^{***}$ \\ 
  & (0.036) & (0.012) & (0.022) & (0.000) & (0.029) & (0.002) & (0.001) \\ 
  & & & & & & & \\ 
 distance\_betweenness & $-$0.337$^{***}$ & $-$0.196$^{***}$ & $-$0.094$^{***}$ & $-$0.002$^{***}$ & $-$0.403$^{***}$ & $-$0.007$^{**}$ & $-$0.006$^{***}$ \\ 
  & (0.046) & (0.012) & (0.028) & (0.000) & (0.026) & (0.003) & (0.001) \\ 
  & & & & & & & \\ 
 area\_difference:ellipticity & 0.094$^{**}$ & 0.067$^{***}$ & 0.023 & 0.001$^{***}$ & $-$0.065$^{**}$ & $-$0.000 & 0.001 \\ 
  & (0.039) & (0.013) & (0.016) & (0.000) & (0.032) & (0.002) & (0.001) \\ 
  & & & & & & & \\ 
 ellipticity:distance\_betweenness & 0.041 & $-$0.023$^{**}$ & $-$0.041$^{*}$ & 0.002$^{***}$ & 0.069$^{***}$ & $-$0.006$^{***}$ & $-$0.002$^{***}$ \\ 
  & (0.039) & (0.009) & (0.024) & (0.000) & (0.023) & (0.002) & (0.001) \\ 
  & & & & & & & \\ 
 Constant & 8.360$^{***}$ & 8.640$^{***}$ & 8.791$^{***}$ & 0.310$^{***}$ & 32.676$^{***}$ & $-$0.004 & 0.140$^{***}$ \\ 
  & (0.034) & (0.012) & (0.024) & (0.000) & (0.031) & (0.002) & (0.001) \\ 
  & & & & & & & \\ 
Observations & 567 & 2,607 & 861 & 561 & 2,607 & 861 & 863 \\ 
R$^{2}$ & 0.354 & 0.468 & 0.643 & 1.000 & 0.694 & 0.089 & 0.999 \\ 
\hline \\[-1.8ex] 
\textit{Notes:} & \multicolumn{7}{l}{$^{***}$Significant at the 1 percent level.} \\ 
 & \multicolumn{7}{l}{$^{**}$Significant at the 5 percent level.} \\ 
 & \multicolumn{7}{l}{$^{*}$Significant at the 10 percent level.} \\ 
\end{tabular} 
\end{sidewaystable} 

}

\printbibliography[title=Supplementary References]
\end{refsection}

\end{document}